%% file: main.tex
\documentclass[manuscript,nonacm]{acmart}

\usepackage{preamble}
\input{common.tex}

\acmJournal{TOSEM}
\setcopyright{none}
\begin{document}
\title[Two Views, One Fix: GraphT5 for Vulnerability Repair]{Two Views, One Fix: Multi-Modal Learning for Automated Vulnerability Repair}

\author{Noah Frontzkowski}
\orcid{0009-0008-0409-942X}
\email{noah.frontzkowski@tuhh.de}
\affiliation{
  \institution{Hamburg University of Technology}
  \city{Hamburg}
  \country{Germany}}

\author{Emanuele Iannone}
\orcid{0000-0001-7489-9969}
\email{emanuele.iannone@tuhh.de}
\affiliation{
  \institution{Hamburg University of Technology}
  \city{Hamburg}
  \country{Germany}}

\author{Riccardo Scandariato}
\orcid{0000-0003-3591-7671}
\email{riccardo.scandariato@tuhh.de}
\affiliation{
  \institution{Hamburg University of Technology}
  \city{Hamburg}
  \country{Germany}}

\renewcommand{\shortauthors}{Frontzkowski et al.}

\begin{abstract}
  Many vulnerabilities require edits in non-contiguous code regions that are connected by control- and data-flow dependencies. Token sequences represent these dependencies poorly. In a comparison of \num{24} source-level approaches, those that use code structure serialize it into tokens, use it to guide attention, or use it to select prompt content. None passes the graph itself to the repair model.

  We present \GraphTFive, a graph-augmented Transformer for Automated Vulnerability Repair. \GraphTFive reads two views of a vulnerable function: its source code and a Vulnerability Dependence Graph (VDG), a slice of the Code Property Graph that keeps the dependencies relevant to repair. A \CodeTFive-initialized text encoder and a VDG encoder feed a decoder with gated dual cross-attention.

  We evaluate \GraphTFive on \num{2411} C/C++ test samples from three datasets. It outperforms \VulMaster by an average relative improvement of \pct{17.1} in Exact Match, \pct{19.6} in \BLEUw, and \pct{8.1} in \CodeBLEU, and exceeds zero-shot \GPTFourO and \GeminiTwoFivePro in static similarity on two datasets.

  Under dynamic validation, \GraphTFive produces plausible patches for \num{12} of \num{81} \datasetname{Vul4C} vulnerabilities, tied with the best baseline, and test-passing patches for \num{15} of \num{35} \vulfourj cases, the third-highest reported total.%
\end{abstract}


\maketitle

\section{Introduction}\label{sec:introduction}
\input{sections/introduction}

\section{Pre-Study: Automated Vulnerability Repair Approaches}
\label{sec:prestudy-avr}
\input{sections/prestudy.tex}

\section{\GraphTFive}\label{sec:grapht5}

\subsection{Overview}
\label{sec:grapht5-overview}
\input{sections/overview.tex}

\subsection{The Vulnerability Dependence Graph}\label{sec:the-vulnerability-dependence-graph}
\input{sections/vdg.tex}

\subsection{\GraphTFive Architecture}\label{sec:grapht5-architecture}
\input{sections/grapht5.tex}

\section{Evaluation}
\label{sec:evaluation}
\input{sections/evaluation.tex}

\section{Discussion}\label{sec:discussion}
\input{sections/discussion.tex}

\section{Threats to Validity}\label{sec:validity}
\input{sections/validity.tex}

\section{Related Work}
\input{sections/related_work.tex}

\section{Conclusion}\label{cha:closing}
\input{sections/conclusion.tex}

\section*{Data Availability}\label{sec:data-availability}
\input{sections/data_availability.tex}

\begin{acks}

  This work was partially supported by the EU-funded project Sec4AI4Sec (grant no. 101120393).
\end{acks}

\appendix

\clearpage
\section{Controlled Vocabulary for the AVR Pre-Study}
\label{app:prestudy-vocabulary}
\input{sections/appendix_prestudy_vocabulary.tex}

\clearpage
\section{Prompt Template for Closed-LLM Evaluation}
\label{app:llm-prompt}
\input{sections/appendix_llm_prompt.tex}

\clearpage
\section{Per-CWE \texorpdfstring{\BLEUw and \CodeBLEU}{BLEU-4 and CodeBLEU} Scores}
\label{app:per-cwe-similarity}
\input{sections/appendix_per_cwe_similarity.tex}

\clearpage
\section{\texorpdfstring{\ac{vdg}}{VDG} Excerpt for the Case Study}
\label{app:case-study-vdg}
\input{sections/appendix_case_study_vdg.tex}

\bibliographystyle{ACM-Reference-Format}
\bibliography{references}

\end{document}

%% file: common.tex
\newboolean{showcomments}
\setboolean{showcomments}{true}
\newboolean{peerreview}
\setboolean{peerreview}{true}

\definecolor{gray75}{gray}{.25}
\definecolor{gray50}{gray}{.5}
\definecolor{gray40}{gray}{.6}
\definecolor{gray30}{gray}{.7}
\definecolor{gray20}{gray}{.8}
\definecolor{gray15}{gray}{.85}
\definecolor{gray10}{gray}{.9}
\definecolor{gray05}{gray}{.95}
\definecolor{redbg}{HTML}{F2968F}
\definecolor{greenbg}{HTML}{CDE4AE}
\definecolor{goalcolor}{HTML}{fffff2}
\definecolor{rqboxcolor}{HTML}{f2f2ff}
\definecolor{rqanswercolor}{HTML}{faf9f5}
\definecolor{takeawaycolor}{HTML}{f2fff2}
\definecolor{pastelblue}{HTML}{699FDD}
\definecolor{pastelorange}{HTML}{F4A261}
\definecolor{pastelyellow}{HTML}{E9C46A}
\definecolor{pastelred}{HTML}{FF7070}
\definecolor{pastelgreen}{HTML}{C1E1C1}
\colorlet{lightyellow}{yellow!60}

\newtcolorbox{rqbox}[2][]{%
	enhanced jigsaw,
	halign=left, valign=top,
	title=#2,
	colback=lightgray!20,
	colframe=lightgray,
	fonttitle=\scshape,
	coltitle=black,
	colbacktitle=lightgray!10,
	attach boxed title to top left={xshift=1cm,yshift*=1mm-\tcboxedtitleheight},
	varwidth boxed title*=-3cm,
	boxed title style={frame code={
					\path[fill=tcbcolback!70!black]
					([yshift=-1mm,xshift=-1mm]frame.north west)
					arc[start angle=0,end angle=180,radius=1mm]
						([yshift=-1mm,xshift=1mm]frame.north east)
					arc[start angle=180,end angle=0,radius=1mm];
					\path[left color=tcbcolback!80!black,right color=tcbcolback!90!black,
						middle color=tcbcolback!90!black]
					([xshift=-2mm]frame.north west) -- ([xshift=2mm]frame.north east)
						[rounded corners=1mm]-- ([xshift=1mm,yshift=-1mm]frame.north east)
						-- (frame.south east) -- (frame.south west)
						-- ([xshift=-1mm,yshift=-1mm]frame.north west)
						[sharp corners]-- cycle;
				},interior engine=empty,
		},
	width=\textwidth,
	boxsep=4pt,
	right=0pt, left=1pt, top=2pt, bottom=0pt,
	boxrule=0.5pt,
	#1
}

\newcommand{\nb}[3]{
	\fbox{\bfseries\sffamily\scriptsize\color{#3}#1}
	{\sf\small$\blacktriangleright${\color{#3}\textit{\textbf{#2}}}$\blacktriangleleft$}
}

\ifthenelse{\boolean{showcomments}}
{
	\newcommand\EMANUELE[1]{\nb{EI}{#1}{orange}}
	\newcommand\NOAH[1]{\nb{NF}{#1}{cyan}}
	\newcommand\RICCARDO[1]{\nb{RS}{#1}{magenta}}
	\newcommand\NF[1]{\NOAH{#1}}
	\newcommand\EI[1]{\EMANUELE{#1}}
	\newcommand\RS[1]{\RICCARDO{#1}}

	\newcommand\REMOVE[1]{\textcolor{gray20}{#1}}

}
{\newcommand\EMANUELE[1]{}
	\newcommand\NOAH[1]{}
	\newcommand\RICCARDO[1]{}
	\newcommand\EI[1]{}
	\newcommand\NF[1]{}
	\newcommand\RS[1]{}

	\newcommand\REMOVE[1]{}
}

\newcommand\rqOne{\textbf{RQ$_1$}\xspace}
\newcommand\rqTwo{\textbf{RQ$_2$}\xspace}
\newcommand\rqThree{\textbf{RQ$_3$}\xspace}
\newcommand\rqFour{\textbf{RQ$_4$}\xspace}
\newcommand\rqFive{\textbf{RQ$_5$}\xspace}

\newcommand{\cpp}[1]{\mintinline{cpp}{#1}}

\newcommand{\CWE}[1]{\textsc{CWE-#1}}

\newcommand{\modelname}[1]{\textsf{#1}}
\newcommand{\datasetname}[1]{\textup{#1}}
\newcommand{\modelid}[1]{\texttt{#1}}

\newcommand{\Joern}{\modelname{Joern}\xspace}

\newcommand{\CodeBERT}{\modelname{CodeBERT}\xspace}

\newcommand{\CodeTFive}{\modelname{CodeT5}\xspace}
\newcommand{\CodeTFiveLarge}{\modelname{CodeT5-large}\xspace}
\newcommand{\TFive}{\modelname{T5}\xspace}

\newcommand{\VulMaster}{\modelname{VulMaster}\xspace}
\newcommand{\VulMasterLarge}{\modelname{VulMaster-large}\xspace}
\newcommand{\GraphTFive}{\modelname{GraphT5}\xspace}
\newcommand{\VulTerminator}{\modelname{VulTerminator}\xspace}
\newcommand{\VulAdvisor}{\modelname{VulAdvisor}\xspace}
\newcommand{\VRepair}{\modelname{VRepair}\xspace}
\newcommand{\VulRepair}{\modelname{VulRepair}\xspace}
\newcommand{\VulRep}{\modelname{VulRep}\xspace}
\newcommand{\VulRTwo}{\modelname{Vul-R2}\xspace}
\newcommand{\VulnFix}{\modelname{VulnFix}\xspace}
\newcommand{\VQM}{\modelname{VQM}\xspace}
\newcommand{\SPVF}{\modelname{SPVF}\xspace}
\newcommand{\SPVR}{\modelname{SPVR}\xspace}

\newcommand{\VuRLE}{\modelname{VuRLE}\xspace}
\newcommand{\SeqTrans}{\modelname{SeqTrans}\xspace}
\newcommand{\Senx}{\modelname{Senx}\xspace}
\newcommand{\ExtractFix}{\modelname{ExtractFix}\xspace}
\newcommand{\VRpilot}{\modelname{VRpilot}\xspace}
\newcommand{\APPATCH}{\modelname{APPATCH}\xspace}
\newcommand{\SanTwoPatch}{\modelname{San2Patch}\xspace}
\newcommand{\PatchAgent}{\modelname{PatchAgent}\xspace}
\newcommand{\CodeRoverS}{\modelname{CodeRover-S}\xspace}
\newcommand{\LoopRepair}{\modelname{LoopRepair}\xspace}
\newcommand{\NTR}{\modelname{NTR}\xspace}
\newcommand{\GumTree}{\modelname{GumTree}\xspace}
\newcommand{\SySeVR}{\modelname{SySeVR}\xspace}
\newcommand{\CodeTwoVec}{\modelname{code2vec}\xspace}
\newcommand{\CodeTwoSeq}{\modelname{code2seq}\xspace}
\newcommand{\Devign}{\modelname{Devign}\xspace}
\newcommand{\FUNDED}{\modelname{FUNDED}\xspace}
\newcommand{\Hoppity}{\modelname{Hoppity}\xspace}
\newcommand{\SlicedLocator}{\modelname{SlicedLocator}\xspace}
\newcommand{\VulLMGNN}{\modelname{Vul-LMGNN}\xspace}
\newcommand{\TreeGen}{\modelname{TreeGen}\xspace}
\newcommand{\Graphormer}{\modelname{Graphormer}\xspace}
\newcommand{\StructCoder}{\modelname{StructCoder}\xspace}
\newcommand{\CSATrans}{\modelname{CSA-Trans}\xspace}
\newcommand{\Codex}{\modelname{Codex}\xspace}
\newcommand{\Incoder}{\modelname{InCoder}\xspace}
\newcommand{\CodeGEN}{\modelname{CodeGen}\xspace}
\newcommand{\GPTFourO}{\modelname{GPT-4o}\xspace}
\newcommand{\GeminiTwoFivePro}{\modelname{Gemini 2.5 Pro}\xspace}

\newcommand{\CodeBLEU}{\textsc{CodeBLEU}\xspace}
\newcommand{\EM}{\textsc{EM}\xspace}

\newcommand{\BLEUw}{\textsc{BLEU-4}\xspace}
\newcommand{\EMw}{\textsc{EM}\textsubscript{weighted}\xspace}
\newcommand{\CodeBERTScore}{\textsc{CodeBERTScore}\xspace}

\newcommand{\BigVul}{\datasetname{Big-Vul}\xspace}
\newcommand{\CVEfixes}{\datasetname{CVEfixes}\xspace}
\newcommand{\ICVul}{\datasetname{ICVul}\xspace}
\newcommand{\PrimeVul}{\datasetname{PrimeVul}\xspace}

\newcommand{\REEF}{\datasetname{REEF}\xspace}
\newcommand{\CodeSearchNet}{\datasetname{CodeSearchNet}\xspace}
\newcommand{\vulfourj}{\datasetname{Vul4J}\xspace}
\newcommand{\vulfourjplus}{\datasetname{Vul4J+}\xspace}
\newcommand{\vulfourjl}{\datasetname{Vul4JL}\xspace}

\newcommand{\pct}[1]{\SI{#1}{\percent}}

%% file: sections/introduction.tex
The number of disclosed software vulnerabilities keeps growing, with thousands of new \acp{cve} each year~\cite{nationalvulnerabilitydatabase:2026:statistics}. Manual auditing and patching is slow, error-prone, and does not scale with modern software development~\cite{hu:2025:sok}. \ac{avr} addresses this problem by generating repairs directly from vulnerable code.

Early attempts to apply general \ac{apr} tools to vulnerabilities had limited success~\cite{bui:emse2024:apr4vul}. Subsequent work designed models specifically for vulnerabilities, including \VRepair~\cite{chen:tse2023:neural}, \VulRepair~\cite{fu:fse2022:vulrepair}, and \VulMaster~\cite{zhou:icse2024:out}. These models frame vulnerability repair as sequence generation, with source code as the main input.

Source text alone, however, does not make explicit the relations that make code vulnerable. Many vulnerabilities depend on control- and data-flow relations between distant parts of the token stream~\cite{yamaguchi:sp2014:modeling,li:tdsc2022:sysevr,wang:tifs2020:combining}. For example, a bounds check may guard a later memory access, and a value may become unsafe only after it propagates through several statements. A repair model should therefore reason over both the code as written and the dependencies that explain how the vulnerability arises.

A common strategy is to extract a tree or graph from the code and linearize it into a token sequence. \ac{sbt}, for example, serializes an \ac{ast} with brackets so that the tree can be recovered from the sequence. Such linearizations can help code transformation tasks~\cite{mahbub:icse2023:explaining, hu:icpc2018:deep, zhou:icse2024:out}, and they leave the model architecture unchanged. Path-based approaches instead represent code through selected \ac{ast} paths: \CodeTwoVec represents a method as a set of path contexts, and \CodeTwoSeq encodes paths as short sequences that the decoder selects through attention~\cite{alon:icse2019:code2vec,alon:iclr2018:code2seq}. In both cases, the model no longer sees the connectivity of the original graph.

Graph neural models avoid this loss because they operate on the graph itself. In software security, however, they have mostly been used for vulnerability detection and classification~\cite{wang:tifs2020:combining, chakraborty:tse2021:deep, wu:cose2023:slicedlocator}, which output a label or a score rather than a repair. Current repair models therefore either use source text alone or use structure only after converting it back into text.

We present \GraphTFive, an \ac{avr} model that combines source code with a vulnerability-centered graph. A dedicated graph encoder propagates information along the dependency relations of the code, and gated dual cross-attention combines both representations during repair generation.

\GraphTFive builds on \ac{t5} and is initialized with pretrained \CodeTFive weights~\cite{wang:emnlp2021:codet5}. It uses one encoder for the source-code passages, one encoder for the \ac{vdg}, and one decoder that attends to both. Each representation therefore keeps its native form instead of being merged into a single token sequence. To the best of our knowledge, \GraphTFive is the first \ac{avr} model to consume a vulnerability-focused graph natively for patch generation, rather than linearizing it into text before learning.

Our evaluation focuses on C/C++, where vulnerabilities often involve explicit memory management, pointer and buffer handling, and low-level control- and data-flow dependencies. We evaluate \GraphTFive on three C/C++ datasets with \num{2411} test samples in total: \BigVul + \CVEfixes~\cite{zhou:icse2024:out}, \ICVul~\cite{lu:2025:icvul}, and \PrimeVul~\cite{ding:icse2025:vulnerability}. Compared with \VulMaster~\cite{zhou:icse2024:out}, \GraphTFive improves Exact Match (\EM) by \pct{10.2} to \pct{21.1} in relative terms, \BLEUw by \pct{13.2} to \pct{24.9}, and \CodeBLEU by \pct{6.3} to \pct{10.5}.
It also achieves higher static similarity scores than zero-shot \GPTFourO and \GeminiTwoFivePro on \BigVul + \CVEfixes and \ICVul, and the highest \BLEUw and \CodeBLEU on \PrimeVul.
With exact matches weighted by the number of repair hunks, \GraphTFive achieves an \EMw of \pct{13.54} compared with \pct{11.91} for \VulMaster, a \pct{13.74} relative gain.
In a separate dynamic validation on the C/C++ benchmark \datasetname{Vul4C}, which validates patches with \ac{poc} exploits and regression tests, \GraphTFive produces plausible patches for \num{12} of \num{81} vulnerabilities, tied with \VulRepair for the best result.
We also adapt the architecture to Java. Starting from the C/C++ checkpoint, we fine-tune \GraphTFive on \num{5295} Java training examples from the vulnerability-fix corpus of Bui et al.~\cite{bui:saner2026:vulterminator}. On the \vulfourj Java benchmark~\cite{bui:msr2022:vul4j}, \GraphTFive produces test-passing patches for \num{15} of \num{35} cases.

In summary, this paper makes the following contributions:
\begin{itemize}[leftmargin=0.5cm]
	\item We introduce \textbf{\aclp{vdg}} (\acsp{vdg}), a vulnerability-focused graph representation derived from \acp{cpg} through slicing. The repair model consumes \acp{vdg} as graphs rather than as token sequences.
	\item We introduce \textbf{\GraphTFive}, a graph-augmented Transformer that combines source code and \acp{vdg} through two encoders and a shared decoder with gated dual cross-attention.
	\item We conduct a \textbf{structured comparison} of \num{24} source-level \ac{avr} approaches by repair paradigm, input representation, localization assumption, and evaluation oracle. It shows that approaches using code structure discard the graph before it reaches the repair model.
	\item We provide a \textbf{replication package} with the implementation, datasets, preprocessing scripts, and evaluation pipeline, available on \href{https://github.com/tuhh-softsec/GraphT5}{GitHub}.
\end{itemize}

%% file: sections/prestudy.tex
\definecolor{prestudyHeader}{HTML}{DDE7F2}
\definecolor{prestudySubheader}{HTML}{EEF3F8}
\definecolor{prestudyStripe}{HTML}{F7F9FB}
\definecolor{prestudyGroup}{HTML}{E8EEF5}
\definecolor{prestudyThisWork}{HTML}{FFF4CC}

\newcommand{\avrYes}{\textcolor{green!45!black}{\ding{51}}}
\newcommand{\avrPart}{\textcolor{orange!80!black}{\ding{108}}}
\newcommand{\avrNo}{\textcolor{red!65!black}{\ding{55}}}
\newcommand{\avrNR}{\textcolor{black!40}{\textsf{N/R}}}
\newcommand{\avrNA}{\textcolor{black!40}{\textsf{N/A}}}
\newcommand{\keygroup}[1]{\par\noindent\hangindent=6.4em\hangafter=1\makebox[6.4em][l]{\textsc{#1}}\ignorespaces}

\ac{avr} takes vulnerable source code as input and generates a patch that removes the vulnerability while preserving the intended functionality. Hu et al.~\cite{hu:2025:sok} describe this workflow in three steps: vulnerability analysis, patch generation, and patch validation. Vulnerability analysis locates and characterizes the vulnerability. Patch validation is either static, without executing the patched program, or dynamic, by executing it with exploits and tests. Before presenting \GraphTFive, we review how prior source-level \ac{avr} approaches implement these steps.

Table~\ref{tab:prestudy-avr} compares studies that (i) are peer-reviewed, (ii) automatically generate source-code patches or actionable repair guidance for software vulnerabilities, and (iii) address general vulnerability classes rather than a single weakness class, defect family, or API domain. Criterion (iii) excludes, for example, cross-site-scripting repair through security unit tests~\cite{mohammadi:issrew2019:automated}.
We collected candidates from the systematizations of knowledge by Hu et al.~\cite{hu:2025:sok} and Li et al.~\cite{li:uss2025:sok} and extended the set through backward and forward snowballing.
We excluded three further groups: (i) security-patch backporting, which transfers an existing fix to other versions of the same software instead of generating a new one; (ii) repair of artifacts other than general-purpose source code, such as binaries and smart contracts; and (iii) generic \ac{apr} systems without a vulnerability-specific design, such as \Hoppity~\cite{dinella:iclr2020:hoppity}. Pinconschi et al.~\cite{pinconschi:issre2021:comparative} and Bui et al.~\cite{bui:emse2024:apr4vul} evaluate such generic systems on vulnerability benchmarks.
We also excluded empirical studies that benchmark existing \ac{avr} approaches without proposing a new one~\cite{wu:issta2023:how,zhang:tdsc2024:pretrained,fu:apsec2023:chatgpt,ságodi:ease2024:reality,han:icse2026:rethinking,wang:tosem2026:evaluation,yadav:acsac2024:rr}.

The final set contains \num{24} approaches. Table~\ref{tab:prestudy-avr} lists them together with \GraphTFive, grouped into four repair paradigms: (1) template- and rule-guided systems, (2) semantics-based approaches driven by program analysis, (3) learning-based neural models, and (4) prompt-based or agentic \ac{llm} pipelines that keep the underlying model frozen.
Across these paradigms, we observe four common limitations.
First, most approaches assume that the vulnerable location is known or can be derived from runtime evidence of the vulnerability, such as a crashing input or a sanitizer report: \num{23} of the \num{24} approaches are evaluated with the vulnerable lines or function given, or with a location derived from an exploit or sanitizer trace.
Second, semantics-based and agentic approaches validate patches dynamically, but they require a \ac{poc} exploit, that is, an input that triggers the vulnerability, or a sanitizer report, together with an environment that can replay it. The same exploit or report also tells them where the vulnerability is, which makes localization much easier than without runtime evidence. Their evaluations are therefore limited to benchmarks with reproducible exploits, which range from a few dozen curated vulnerabilities to samples generated by fuzzing.
Third, learning-based \ac{avr} mostly frames repair as sequence generation over localized code snippets or functions. Large-scale C/C++ evaluations rely mainly on static similarity metrics, such as \EM, \BLEUw, and \CodeBLEU. The reason is that \ac{pov} tests, which fail on the vulnerable code and pass once the vulnerability is fixed (also called vulnerability-witnessing tests), are not available at that scale. Even where exploits or \ac{pov} tests exist, running them requires building each project at its historical vulnerable commit. This is time-consuming and often fails, for example because of broken or unavailable dependencies. \vulfourj retains \num{79} of \num{912} analyzed Java vulnerabilities~\cite{bui:msr2022:vul4j}, and only \num{68} of these \num{79} compiled in the evaluation of Hu et al.~\cite{hu:2025:sok}. For \datasetname{Vul4C}, Hu et al. discarded \num{34} of \num{178} C/C++ vulnerabilities whose exploit did not succeed on the rebuilt program~\cite{hu:2025:sok}.
Fourth, approaches that use structural information serialize the tree or graph into tokens, use it to guide attention, or use it to select prompt content. None of them passes the vulnerability-relevant graph itself to the repair model.

Six approaches in Table~\ref{tab:prestudy-avr} use tree- or graph-derived structure, and all of them discard the tree or graph before it reaches the neural model.
\SPVF~\cite{zhou:emse2022:spvf} and \VulRep~\cite{wei:ejwcn2023:vulrep} flatten a function's \ac{ast} into a token sequence. \VulRep additionally replaces identifiers, literals, and method names with placeholders to limit the vocabulary. They feed the sequence to a pointer-generator network or a standard Transformer, which can recover tree adjacency only implicitly through self-attention, not by passing messages along tree edges.
\VulMaster~\cite{zhou:icse2024:out} follows the same approach. Based on the structure-preserving traversal of Hu et al.~\cite{hu:icpc2018:deep}, it concatenates the type and value of each \ac{ast} node in depth-first order into an ``AST node sequence'', splits the result into \num{512}-token segments, and encodes each segment as a separate \ac{fid} passage. \ac{fid} does not restore the discarded tree edges. It only allows the model to read more tokens than a single encoder pass.
\SeqTrans~\cite{chi:tse2022:seqtrans} uses the \ac{ast} differently: it traverses the tree with a modified \GumTree diff to extract def-use dependencies, but only to select which statements to keep. The kept statements are serialized into a token sequence for a standard \ac{s2s} model, so the dependency graph serves only as a preprocessing filter.
\APPATCH~\cite{nong:uss2025:appatch} and \SPVR~\cite{wang:ase2026:spvr} use structure only outside the neural model. \APPATCH applies union slicing to a \ac{sdg} to select the lines included in the prompt of a frozen \ac{llm}. \SPVR mines repair rules offline from a \ac{met} computed over pairs of vulnerable and fixed \acp{ast} and uses these rules to select the natural-language guidance in the prompt.
In both cases, the graph only determines which text the model sees.

\GraphTFive instead passes its \ac{vdg} to a \textbf{dedicated graph encoder} that performs \textbf{message passing along the graph's edges}, with attention conditioned on edge type and distance (Section~\ref{sec:grapht5-architecture}). The graph encoder is separate from the text encoder, and the two are \textbf{fused only in the decoder} through gated dual cross-attention.
The graph structure therefore constrains attention at every layer of the graph encoder, instead of being recovered from a flattened token sequence or used only to select prompt text.

The studies also differ in evaluation scale and oracle. \VulMaster reports a C/C++ similarity evaluation on \num{1612} samples and a separate Java study on \num{35} cases, and \VQM tests on \num{1090} C/C++ samples. \SeqTrans reports \num{2130} samples from ten-fold cross-validation on one Java dataset. In their original publications, none of the learning-based neural approaches validate their C/C++ repairs by executing tests or exploits. \VulMaster and \SeqTrans use tests only for Java. Hu et al.~\cite{hu:2025:sok} later validated \VRepair, \VulRepair, \VQM, and \VulMaster on \datasetname{Vul4C}, and \rqFour reuses their results. Our evaluation uses \num{2411} C/C++ test samples from three datasets for the static evaluation, and \textbf{dynamic validation on \num{81} C/C++ and \num{35} Java vulnerabilities}.

\begin{rqbox}{Implications for \GraphTFive}
	Prior source-level \ac{avr} approaches either depend on runtime evidence, such as exploits or sanitizer reports, or treat code as text. Approaches that use code structure discard the graph before it reaches the repair model. \GraphTFive addresses these gaps: it \textbf{requires neither an exploit nor a test suite at repair time}, \textbf{gives the repair model the vulnerability-relevant graph as a native input}, alongside the source text, and is validated both statically and dynamically on C/C++ and Java.
\end{rqbox}

\input{sections/prestudy_table}

%% file: sections/prestudy_table.tex
\begin{landscape}
    \begingroup
    \singlespacing
    \fontsize{7}{7.5}\selectfont
    \begin{longtblr}[
            caption = {Pre-study comparison of source-level \ac{avr} approaches, grouped by repair paradigm. Cells use the controlled vocabulary defined in Appendix Table~\ref{tab:prestudy-vocab}.},
            label = {tab:prestudy-avr},
        ]{
            rowhead = 2,
            width = \linewidth,
            colspec = {
                    X[2.1,l,m]
                    X[1.1,c,m]
                    X[1.3,l,m]
                    X[3.3,l,m]
                    X[2.0,l,m]
                    X[1.45,l,m]
                    X[0.8,l,m]
                    X[1.5,l,m]
                    X[1.9,l,m]
                    X[2.9,l,m]
                    X[1.9,l,m]
                    X[1.45,l,m]
                },
            row{1} = {font=\bfseries, bg=prestudyHeader},
            row{2} = {font=\bfseries, bg=prestudySubheader},
            row{3,6,10,23} = {font=\bfseries, bg=prestudyGroup},
            row{4,7,9,11,13,15,17,19,21,24,26,28,30} = {bg=prestudyStripe},
            row{22} = {bg=prestudyThisWork},
            hlines = {white},
            vline{3,6,7,9,12} = {0.45pt, gray!35},
            hline{1,3,32} = {0.8pt, black},
            hline{6,10,23} = {0.45pt, gray!45},
            colsep = 1.2pt,
            rowsep = 0.1pt,
        }
        \SetCell[r=2]{c} Approach                                                                                                                                                                     &
        \SetCell[r=2]{c} Venue                                                                                                                                                                        &
        \SetCell[r=2]{c} {Approach\newline{}/\ Arch.}                                                                                                                                                 &
        \SetCell[r=2]{c} {Model backbone}                                                                                                                                                             &
        \SetCell[r=2]{c} {Input\newline{}representation}                                                                                                                                              &
        \SetCell[r=2]{c} Languages                                                                                                                                                                    &
        \SetCell[c=2]{c} Repair setup                                                                                                                                                                 &
                                                                                                                                                                                                      &
        \SetCell[c=3]{c} Datasets                                                                                                                                                                     &
                                                                                                                                                                                                      &                         &
        \SetCell[r=2]{c} {Evaluation\newline{}oracle}                                                                                                                                                                                                                                 \\
                                                                                                                                                                                                      &                         &                         &   &  &  &
        Loc.                                                                                                                                                                                          & {Repair\newline{}scope} &
        Pretraining                                                                                                                                                                                   & Fine-tuning             & {Testing\newline{}size} &                           \\
        \SetCell[c=12]{l} Template- and rule-guided approaches                                                                                                                                        &                         &                         &   &  &  &   &  &  &  &  & \\
        \VuRLE~\cite{ma:esorics2017:vurle}                                                                                                                                                            &
        {ESORICS\newline{}2017}                                                                                                                                                                       &
        PAT                                                                                                                                                                                           &
        \avrNA                                                                                                                                                                                        &
        TXT                                                                                                                                                                                           &
        Java                                                                                                                                                                                          &
        E2E                                                                                                                                                                                           &
        File                                                                                                                                                                                          &
        \avrNA                                                                                                                                                                                        &
        \datasetname{VuRLE Java examples}                                                                                                                                                             &
        Java: \num{279}                                                                                                                                                                               &
        MAN                                                                                                                                                                                                                                                                           \\
        \VulTerminator~\cite{bui:saner2026:vulterminator}                                                                                                                                             &
        {SANER\newline{}2026}                                                                                                                                                                         &
        TPL + MLM                                                                                                                                                                                     &
        \modelname{UniXCoder}                                                                                                                                                                         &
        {TXT + LOC\newline{}+ CWE}                                                                                                                                                                    &
        Java                                                                                                                                                                                          &
        PFL-L                                                                                                                                                                                         &
        Single/multi-hunk                                                                                                                                                                             &
        {\CodeSearchNet;\newline{}\datasetname{C4}}                                                                                                                                                   &
        \datasetname{Bui Java fixes} (\num{4147})                                                                                                                                                     &
        Java: \num{275}                                                                                                                                                                               &
        {COMP, PoV,\newline{}TEST, MAN}                                                                                                                                                                                                                                               \\
        \SetCell[c=12]{l} Semantics-based approaches                                                                                                                                                  &                         &                         &   &  &  &   &  &  &  &  & \\
        \Senx~\cite{huang:sp2019:using}                                                                                                                                                               &
        {S\&P\newline{}2019}                                                                                                                                                                          &
        {SYMB\newline{}+ SYNTH}                                                                                                                                                                       &
        \avrNA                                                                                                                                                                                        &
        POC + PROP                                                                                                                                                                                    &
        C/C++                                                                                                                                                                                         &
        PoC-L                                                                                                                                                                                         &
        Cross-function                                                                                                                                                                                &
        \avrNA                                                                                                                                                                                        &
        \avrNA                                                                                                                                                                                        &
        C/C++: \num{42}                                                                                                                                                                               &
        {EXP, TEST,\newline{}MAN}                                                                                                                                                                                                                                                     \\
        \ExtractFix~\cite{gao:2021:tests}                                                                                                                                                             &
        {TOSEM\newline{}2021}                                                                                                                                                                         &
        {SYMB\newline{}+ SYNTH}                                                                                                                                                                       &
        \avrNA                                                                                                                                                                                        &
        POC + SAN                                                                                                                                                                                     &
        C/C++                                                                                                                                                                                         &
        PoC-L                                                                                                                                                                                         &
        Multi-location                                                                                                                                                                                &
        \avrNA                                                                                                                                                                                        &
        \avrNA                                                                                                                                                                                        &
        C/C++: \num{56}                                                                                                                                                                               &
        EXP, MAN                                                                                                                                                                                                                                                                      \\
        \VulnFix~\cite{zhang:issta2022:program}                                                                                                                                                       &
        {ISSTA\newline{}2022}                                                                                                                                                                         &
        INVAR                                                                                                                                                                                         &
        \avrNA                                                                                                                                                                                        &
        {POC + LOC\newline{}+ SAN}                                                                                                                                                                    &
        C/C++                                                                                                                                                                                         &
        PFL-L                                                                                                                                                                                         &
        Statement                                                                                                                                                                                     &
        \avrNA                                                                                                                                                                                        &
        \avrNA                                                                                                                                                                                        &
        C/C++: \num{39}                                                                                                                                                                               &
        MAN                                                                                                                                                                                                                                                                           \\
        \SetCell[c=12]{l} Learning-based neural approaches                                                                                                                                            &                         &                         &   &  &  &   &  &  &  &  & \\
        GAN-based repair~\cite{harer:neurips2018:learning}                                                                                                                                            &
        {NeurIPS\newline{}2018}                                                                                                                                                                       &
        GAN (NMT)                                                                                                                                                                                     &
        {Custom NMT\newline{}+ discriminator}                                                                                                                                                         &
        TXT                                                                                                                                                                                           &
        C/C++                                                                                                                                                                                         &
        PFL-Func                                                                                                                                                                                      &
        Function                                                                                                                                                                                      &
        \datasetname{SATE IV autoencoder}                                                                                                                                                             &
        \datasetname{SATE IV} (synthetic)                                                                                                                                                             &
        C/C++: \num{4117}                                                                                                                                                                             &
        SIM                                                                                                                                                                                                                                                                           \\
        \SPVF~\cite{zhou:emse2022:spvf}                                                                                                                                                               &
        {EMSE\newline{}2022}                                                                                                                                                                          &
        PG/Attn.                                                                                                                                                                                      &
        Custom pointer-generator                                                                                                                                                                      &
        {TXT + AST-SEQ\newline{}+ PROP + CWE}                                                                                                                                                         &
        C/C++; Python                                                                                                                                                                                 &
        PFL-Func                                                                                                                                                                                      &
        Function                                                                                                                                                                                      &
        \avrNA                                                                                                                                                                                        &
        {\datasetname{Fan C/C++} (\num{8591});\newline{}\datasetname{Wartschinski Python} (\num{3624})}                                                                                               &
        {C/C++: \num{1074};\newline{}Python: \num{453}}                                                                                                                                               &
        SIM, COMP                                                                                                                                                                                                                                                                     \\
        \SeqTrans~\cite{chi:tse2022:seqtrans}                                                                                                                                                         &
        {TSE\newline{}2022}                                                                                                                                                                           &
        TR-E/D                                                                                                                                                                                        &
        {Custom TR\newline{}(bug-fix PT)}                                                                                                                                                             &
        TXT + DF                                                                                                                                                                                      &
        Java                                                                                                                                                                                          &
        PFL-L                                                                                                                                                                                         &
        Statement                                                                                                                                                                                     &
        \datasetname{Tufano Java bug fixes}                                                                                                                                                           &
        \datasetname{Ponta Java vulns.}                                                                                                                                                               &
        Java: \num{2130}                                                                                                                                                                              &
        {SIM, SAST,\newline{}COMP, TEST,\newline{}MAN}                                                                                                                                                                                                                                \\
        \VulRepair~\cite{fu:fse2022:vulrepair}                                                                                                                                                        &
        {FSE\newline{}2022}                                                                                                                                                                           &
        TR-E/D                                                                                                                                                                                        &
        \modelname{CodeT5-base}                                                                                                                                                                       &
        {TXT + LOC\newline{}+ CWE}                                                                                                                                                                    &
        C/C++                                                                                                                                                                                         &
        PFL-L                                                                                                                                                                                         &
        Function                                                                                                                                                                                      &
        {\CodeSearchNet;\newline{}\datasetname{C/C\# corpus}}                                                                                                                                         &
        \BigVul{}+\CVEfixes                                                                                                                                                                           &
        C/C++: \num{1706}                                                                                                                                                                             &
        SIM                                                                                                                                                                                                                                                                           \\
        \VRepair~\cite{chen:tse2023:neural}                                                                                                                                                           &
        {TSE\newline{}2023}                                                                                                                                                                           &
        TR-E/D                                                                                                                                                                                        &
        {Custom TR\newline{}(bug-fix PT)}                                                                                                                                                             &
        {TXT + LOC\newline{}+ CWE}                                                                                                                                                                    &
        C                                                                                                                                                                                             &
        PFL-L                                                                                                                                                                                         &
        {Function;\newline{}multi-location}                                                                                                                                                           &
        \datasetname{Chen C bug fixes}                                                                                                                                                                &
        \BigVul; \CVEfixes                                                                                                                                                                            &
        C: \num{1430}                                                                                                                                                                                 &
        SIM                                                                                                                                                                                                                                                                           \\
        \VulRep~\cite{wei:ejwcn2023:vulrep}                                                                                                                                                           &
        {EJWCN\newline{}2023}                                                                                                                                                                         &
        {TR-E/D\newline{}+ HYB}                                                                                                                                                                       &
        Custom TR                                                                                                                                                                                     &
        {TXT + AST-SEQ\newline{}+ VIC}                                                                                                                                                                &
        C/C++; Java                                                                                                                                                                                   &
        PFL-L                                                                                                                                                                                         &
        Function                                                                                                                                                                                      &
        \avrNA                                                                                                                                                                                        &
        \BigVul{} (\num{3754})                                                                                                                                                                        &
        C/C++ + Java: \num{116}                                                                                                                                                                       &
        MAN                                                                                                                                                                                                                                                                           \\
        \VQM~\cite{fu:tosem2024:vision}                                                                                                                                                               &
        {TOSEM\newline{}2024}                                                                                                                                                                         &
        TR-E/D                                                                                                                                                                                        &
        {\CodeTFive +\newline{}\CodeBERT}                                                                                                                                                             &
        {TXT + LOC\newline{}+ VQ}                                                                                                                                                                     &
        C/C++                                                                                                                                                                                         &
        PFL-L                                                                                                                                                                                         &
        Function                                                                                                                                                                                      &
        {\CodeSearchNet;\newline{}\datasetname{C/C\#}; \datasetname{Chen C bug fixes}}                                                                                                                &
        \BigVul{}+\CVEfixes                                                                                                                                                                           &
        C/C++: \num{1090}                                                                                                                                                                             &
        SIM                                                                                                                                                                                                                                                                           \\
        \VulMaster~\cite{zhou:icse2024:out}                                                                                                                                                           &
        {ICSE\newline{}2024}                                                                                                                                                                          &
        TR-E/D + FiD                                                                                                                                                                                  &
        {\modelname{CodeT5-base};\newline{}\modelname{CodeT5p-large} (Java)}                                                                                                                          &
        {TXT + LOC\newline{}+ AST-SEQ\newline{}+ CWE}                                                                                                                                                 &
        C/C++; Java                                                                                                                                                                                   &
        PFL-L                                                                                                                                                                                         &
        Function                                                                                                                                                                                      &
        {\CodeSearchNet;\newline{}\datasetname{C/C\#}; \datasetname{Chen C bug fixes}}                                                                                                                &
        {\BigVul{}+\CVEfixes;\newline{}\datasetname{Zhu Java bug fixes}}                                                                                                                              &
        {C/C++: \num{1612};\newline{}Java: \num{35}}                                                                                                                                                  &
        {SIM, PoV,\newline{}MAN}                                                                                                                                                                                                                                                      \\
        \VulAdvisor~\cite{zhang:ase2024:vuladvisor}                                                                                                                                                   &
        {ASE\newline{}2024}                                                                                                                                                                           &
        TR-E/D                                                                                                                                                                                        &
        \modelname{CodeT5-base}                                                                                                                                                                       &
        TXT                                                                                                                                                                                           &
        C/C++                                                                                                                                                                                         &
        PFL-Func                                                                                                                                                                                      &
        \avrNA                                                                                                                                                                                        &
        {\CodeSearchNet;\newline{}\datasetname{C/C\# corpus}}                                                                                                                                         &
        \datasetname{VulAdvisor} (\num{18517})                                                                                                                                                        &
        C/C++: \num{1852}                                                                                                                                                                             &
        {SIM, MAN}                                                                                                                                                                                                                                                                    \\
        {Fine-tuned code\newline{}LLMs~\cite{de-fitero-dominguez:eaai2024:enhanced}}                                                                                                                  &
        {EAAI\newline{}2024}                                                                                                                                                                          &
        TR-DO (FT)                                                                                                                                                                                    &
        {\modelname{Code Llama 7B};\newline{}\modelname{Mistral 7B}}                                                                                                                                  &
        {TXT + LOC\newline{}+ CWE}                                                                                                                                                                    &
        C/C++                                                                                                                                                                                         &
        PFL-L                                                                                                                                                                                         &
        Function                                                                                                                                                                                      &
        LLM corpora \avrNR                                                                                                                                                                            &
        \BigVul{}+\CVEfixes                                                                                                                                                                           &
        C/C++: \num{1706}                                                                                                                                                                             &
        SIM                                                                                                                                                                                                                                                                           \\
        \VulRTwo~\cite{wen:2025:vulr2}                                                                                                                                                                &
        {ASE\newline{}2025}                                                                                                                                                                           &
        {TR-DO (FT)\newline{}+ RL}                                                                                                                                                                    &
        {\modelname{Qwen2.5-14B-}\newline{}\modelname{Instruct-1M}}                                                                                                                                   &
        {TXT + LOC\newline{}+ CWE + NL}                                                                                                                                                               &
        C/C++                                                                                                                                                                                         &
        PFL-L                                                                                                                                                                                         &
        Function                                                                                                                                                                                      &
        LLM corpora \avrNR                                                                                                                                                                            &
        {\PrimeVul{} reasoning\newline{}data}                                                                                                                                                         &
        C/C++: \num{819}                                                                                                                                                                              &
        SIM                                                                                                                                                                                                                                                                           \\
        \textbf{\GraphTFive}                                                                                                                                                                          &
        {This\newline{}work}                                                                                                                                                                          &
        TR-E/D + GNN                                                                                                                                                                                  &
        \modelname{CodeT5-base}                                                                                                                                                                       &
        {TXT + LOC\newline{}+ CWE + CPG\newline{} + FUSE}                                                                                                                                             &
        C/C++; Java                                                                                                                                                                                   &
        PFL-L                                                                                                                                                                                         &
        {Function;\newline{}single/multi-hunk}                                                                                                                                                        &
        {\CodeSearchNet;\newline{}\datasetname{C/C\#}; \datasetname{Chen C bug fixes}}                                                                                                                &
        {\BigVul{}+\CVEfixes;\newline{}\ICVul; \PrimeVul;\newline{}\datasetname{Bui Java fixes}}                                                                                                      &
        {C/C++: \num{2411};\newline{}\datasetname{Vul4C}: \num{81};\newline{}Java: \num{35}}                                                                                                          &
        {SIM, COMP,\newline{}PoV, EXP,\newline{}TEST}                                                                                                                                                                                                                                 \\
        \pagebreak\SetCell[c=12]{l} Prompt-based and agentic LLM approaches                                                                                                                           &                         &                         &   &  &  &   &  &  &  &  & \\
        {Zero-shot\newline{}LLM repair~\cite{pearce:sp2023:examining}}                                                                                                                                &
        {S\&P\newline{}2023}                                                                                                                                                                          &
        TR-DO                                                                                                                                                                                         &
        {\modelid{code-cushman-001};\newline{}\modelid{code-davinci-001/002};\newline{}\modelname{J1-Large/Jumbo}; \modelname{PolyCoder};\newline{}\modelname{GPT2-Csrc}}                             &
        PROMPT                                                                                                                                                                                        &
        {C/C++;\newline{}Python;\newline{}Verilog}                                                                                                                                                    &
        PFL-L                                                                                                                                                                                         &
        {File;\newline{}single-hunk}                                                                                                                                                                  &
        LLM corpora \avrNR                                                                                                                                                                            &
        \avrNA                                                                                                                                                                                        &
        {C/C++: \num{112};\newline{}Python: \num{24};\newline{}Verilog: \num{2}}                                                                                                                      &
        {SAST, COMP,\newline{}EXP, TEST, MAN}                                                                                                                                                                                                                                         \\
        \VRpilot~\cite{kulsum:aiware2024:case}                                                                                                                                                        &
        {AIware\newline{}2024}                                                                                                                                                                        &
        TR-DO + CoT                                                                                                                                                                                   &
        {\modelname{GPT-3.5 Turbo}\newline{}(version \avrNR)}                                                                                                                                         &
        {TXT + NL\newline{}+ FDBK}                                                                                                                                                                    &
        C; Java                                                                                                                                                                                       &
        PFL-Func                                                                                                                                                                                      &
        Function                                                                                                                                                                                      &
        LLM corpora \avrNR                                                                                                                                                                            &
        \avrNA                                                                                                                                                                                        &
        {C: \num{10};\newline{}Java: \num{50}}                                                                                                                                                        &
        {COMP, PoV, EXP,\newline{}TEST, MAN}                                                                                                                                                                                                                                          \\
        \APPATCH~\cite{nong:uss2025:appatch}                                                                                                                                                          &
        {USENIX\newline{}2025}                                                                                                                                                                        &
        TR-DO + HYB                                                                                                                                                                                   &
        {\modelname{Gemini 1.5 Pro}; \modelname{Claude 3.5 Sonnet};\newline{}\modelname{GPT-4 Turbo}; \modelname{Llama 3.1 70B}}                                                                      &
        {SLICE + PROMPT\newline{}+ ICL}                                                                                                                                                               &
        C                                                                                                                                                                                             &
        {PFL-L/\newline{}E2E}                                                                                                                                                                         &
        Cross-function                                                                                                                                                                                &
        LLM corpora \avrNR                                                                                                                                                                            &
        \avrNA                                                                                                                                                                                        &
        C: \num{117}                                                                                                                                                                                  &
        SIM, MAN                                                                                                                                                                                                                                                                      \\
        \SanTwoPatch~\cite{kim:uss2025:logs}                                                                                                                                                          &
        {USENIX\newline{}2025}                                                                                                                                                                        &
        TR-DO + ToT                                                                                                                                                                                   &
        {\modelid{gpt-4o-2024-08-06}; \modelname{GPT-4o mini};\newline{}\modelname{GPT-3.5 Turbo}; \modelname{Claude 3.5 Sonnet};\newline{}\modelname{Gemini 1.5 Pro/Flash}}                          &
        {TXT + POC\newline{}+ SAN}                                                                                                                                                                    &
        C/C++                                                                                                                                                                                         &
        PoC-L                                                                                                                                                                                         &
        Single/multi-hunk                                                                                                                                                                             &
        LLM corpora \avrNR                                                                                                                                                                            &
        \avrNA                                                                                                                                                                                        &
        C/C++: \num{66}                                                                                                                                                                               &
        {COMP, EXP,\newline{}TEST, MAN}                                                                                                                                                                                                                                               \\
        \PatchAgent~\cite{yu:uss2025:patchagent}                                                                                                                                                      &
        {USENIX\newline{}2025}                                                                                                                                                                        &
        {TR-DO\newline{}+ AGENT}                                                                                                                                                                      &
        {\modelid{gpt-4-0125-preview}; \modelid{gpt-4o-2024-05-13};\newline{}\modelid{claude-3-}\allowbreak\modelid{\{opus,sonnet,haiku\}-}\allowbreak\modelid{20240229}}                             &
        {TXT + POC\newline{}+ SAN}                                                                                                                                                                    &
        C/C++                                                                                                                                                                                         &
        PoC-L                                                                                                                                                                                         &
        Multi-location                                                                                                                                                                                &
        LLM corpora \avrNR                                                                                                                                                                            &
        \avrNA                                                                                                                                                                                        &
        C/C++: \num{178}                                                                                                                                                                              &
        {COMP, EXP,\newline{}TEST}                                                                                                                                                                                                                                                    \\
        \SPVR~\cite{wang:ase2026:spvr}                                                                                                                                                                &
        {AuSE\newline{}2026}                                                                                                                                                                          &
        {TR-DO\newline{}+ HYB}                                                                                                                                                                        &
        {\modelid{gpt-3.5-turbo-0125}; \modelname{GPT-4 Turbo};\newline{}\modelname{Gemini 2.5 Pro}; \modelname{DeepSeek-Coder-6.7B-}\newline{}\modelname{Instruct}; \modelname{Magicoder-S-DS-6.7B}} &
        {SLICE + PROMPT\newline{}+ CWE}                                                                                                                                                               &
        C/C++                                                                                                                                                                                         &
        PFL-L                                                                                                                                                                                         &
        Function                                                                                                                                                                                      &
        LLM corpora \avrNR                                                                                                                                                                            &
        \avrNA                                                                                                                                                                                        &
        C/C++: \num{547}                                                                                                                                                                              &
        SIM                                                                                                                                                                                                                                                                           \\
        \CodeRoverS~\cite{zhang:icse-seip2026:fixing}                                                                                                                                                 &
        {ICSE-SEIP\newline{}2026}                                                                                                                                                                     &
        {TR-DO\newline{}+ AGENT}                                                                                                                                                                      &
        {\modelname{o3-mini};\newline{}\modelname{Gemini 2.5 Flash}}                                                                                                                                  &
        {TXT + POC\newline{}+ SAN}                                                                                                                                                                    &
        C/C++                                                                                                                                                                                         &
        PoC-L                                                                                                                                                                                         &
        \avrNR                                                                                                                                                                                        &
        LLM corpora \avrNR                                                                                                                                                                            &
        \avrNA                                                                                                                                                                                        &
        C/C++: \num{633}                                                                                                                                                                              &
        {COMP, EXP,\newline{}MAN}                                                                                                                                                                                                                                                     \\
        \LoopRepair~\cite{ye:arxiv2026:well}                                                                                                                                                          &
        {ICSE\newline{}2026}                                                                                                                                                                          &
        TR-DO                                                                                                                                                                                         &
        {\modelid{gpt-4o-mini-2024-07-18}; \modelname{GPT-3.5 Turbo};\newline{}\modelname{Claude 3.5 Sonnet}; \modelid{gemini-1.5-flash-002};\newline{}\modelname{DeepSeek-V3}}                       &
        {TXT + LOC\newline{}+ POC + FDBK}                                                                                                                                                             &
        C/C++; Java                                                                                                                                                                                   &
        PoC-L                                                                                                                                                                                         &
        {Function;\newline{}single/multi-hunk}                                                                                                                                                        &
        LLM corpora \avrNR                                                                                                                                                                            &
        \avrNA                                                                                                                                                                                        &
        {C/C++: \num{40};\newline{}Java: \num{5}}                                                                                                                                                     &
        {COMP, PoV,\newline{}MAN}                                                                                                                                                                                                                                                     \\
    \end{longtblr}\par
    \vspace{1pt}
    {\fontsize{7}{8}\selectfont
        \keygroup{General} \textbf{\avrNR}: not reported; \textbf{\avrNA}: not applicable. Appendix Table~\ref{tab:prestudy-vocab} gives the complete definitions.
        \keygroup{Localization} \textbf{PFL-L}: perfect line-level localization; \textbf{PFL-Func}: vulnerable function given; \textbf{E2E}: localization performed by the approach; \textbf{PoC-L}: location derived from a \ac{poc} exploit or sanitizer trace.
        \keygroup{Oracles} \textbf{SIM}: static similarity; \textbf{SAST}: static analysis; \textbf{COMP}: compilation; \textbf{PoV}: \ac{pov} tests; \textbf{EXP}: exploit replay; \textbf{TEST}: regression, unit, or functional tests; \textbf{MAN}: manual validation.
        \keygroup{Datasets} \textbf{\BigVul{}} (\textbf{Fan C/C++})~\cite{fan:msr2020:code}; \textbf{Bui Java fixes}~\cite{bui:saner2026:vulterminator}; \textbf{C/C\# corpus}~\cite{wang:emnlp2021:codet5}; \textbf{C4}~\cite{raffel:2020:exploring}; \textbf{Chen C bug fixes}~\cite{chen:tse2023:neural}; \textbf{\CodeSearchNet}~\cite{husain:2019:codesearchnet}; \textbf{\CVEfixes}~\cite{bhandari:promise2021:cvefixes}; \textbf{\ICVul}~\cite{lu:2025:icvul}; \textbf{Ponta Java vulns.}~\cite{ponta:msr2019:manuallycurated}; \textbf{\PrimeVul}~\cite{ding:icse2025:vulnerability}; \textbf{SATE IV}~\cite{okun:2013:report}; \textbf{Tufano Java bug fixes}~\cite{tufano:2019:empirical}; \textbf{Vul4C}~\cite{hu:2025:sok}; \textbf{Wartschinski Python}~\cite{wartschinski:2022:vudenc}; \textbf{Zhu Java bug fixes}~\cite{zhu:fse2021:syntaxguided}. Datasets without a key entry, such as \datasetname{VuRLE Java examples} and \datasetname{VulAdvisor}, were introduced by the row's own paper, which the \textit{Approach} column cites.\\
        \par\noindent\textbf{Notes.} \textit{Testing size} gives the evaluated samples per language; counts from disjoint datasets are listed separately or summed, and for protocols that reuse one corpus, the largest evaluated set is shown. \textit{Evaluation oracle} lists every reported validation mechanism, including subset studies, so an oracle may cover fewer cases than the testing size: \VulMaster's PoV and MAN refer to its \num{35}-case \vulfourj study, \VulTerminator's PoV and TEST cover only the \num{106} \datasetname{Vul4J+} cases because \datasetname{Vul4JL} has no tests, \CodeRoverS manually inspects \num{88} plausible patches, and \VulAdvisor's human study covers \num{100} of its \num{1852} test samples. \textit{Model backbone} names the checkpoint or API model that generates the repairs; \textit{Custom} marks an architecture trained from scratch. \textit{Pretraining} ``LLM corpora \avrNR'' marks backbones whose pretraining corpus is not publicly documented. The zero-shot \ac{llm} row is the prompting setup of Pearce et al., which uses the developer patch as localization oracle. \VulAdvisor{} generates natural-language guidance rather than patches.\par}
    \endgroup
\end{landscape}

%% file: sections/overview.tex
\begin{figure}[htbp]%
	\centering
	\includegraphics[width=\linewidth]{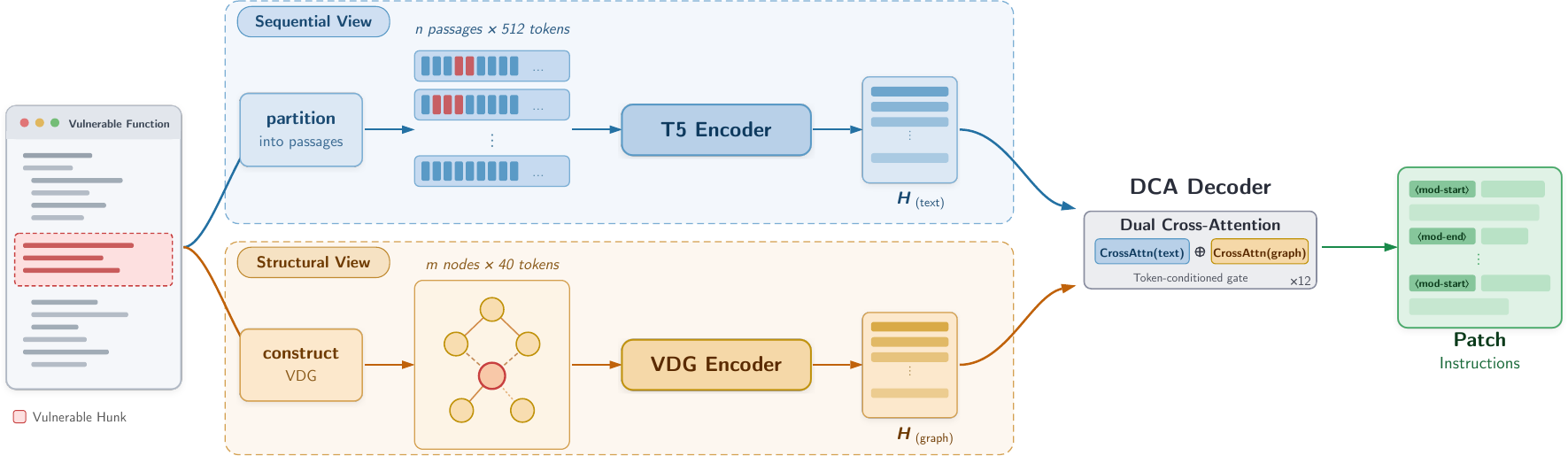}
	\caption{Outline of \GraphTFive's architecture. Each passage spans \num{512} tokens, while each node uses up to \num{40} tokens.}
	\Description{A vulnerable function supplies source-code passages to the text encoder and a vulnerability dependence graph to the graph encoder. Both representations feed the repair decoder.}
	\label{fig:introduction:grapht5-overview}
\end{figure}

\GraphTFive takes two inputs: the source code of the vulnerable function and its \ac{vdg}.
As Figure~\ref{fig:introduction:grapht5-overview} shows, two encoders process these inputs in parallel: a \textit{text encoder}, initialized from \CodeTFive, encodes the code in passages of \num{512} tokens, and a \textit{graph encoder} encodes the \ac{vdg}.
Following the \ac{fid} principle of encoding contexts separately and fusing them in the decoder~\cite{izacard:eacl2021:leveraging}, the decoder combines both encoder outputs through \textit{\ac{dca}}.
The model can therefore use both the sequential code context and the structural dependencies when generating repair instructions.
As in prior \ac{avr} work, we assume that the vulnerable hunks are already localized and marked in the input code~\cite{zhou:emse2022:spvf, chen:tse2023:neural, fu:fse2022:vulrepair, zhou:icse2024:out}. \GraphTFive therefore addresses patch generation only; vulnerability detection and localization are outside the scope of this work.

%% file: sections/vdg.tex
\acp{vdg} address three limitations of \acp{cpg} for repair generation. \textbf{First, edge redundancy}: a \ac{cpg} contains several overlapping edge types, such as \texttt{AST}, \texttt{CFG}, \texttt{CDG}, \texttt{REACHING\_DEF}, and \texttt{REF}. These create many alternative paths between the same code elements, which dilutes the vulnerability-relevant signal. \textbf{Second, structural over-complexity}: \acp{cpg} contain auxiliary constructs, such as \texttt{BLOCK} nodes and type- or binding-related nodes. These support general static analysis but enlarge the graph without helping repair. \textbf{Third, limited source alignment}: some abstract \ac{cpg} nodes do not map to concrete source spans, so the model cannot easily relate graph information to the tokens it must edit.
A \ac{vdg} therefore reduces the \ac{cpg} to a source-aligned graph centered on the vulnerability.

\begin{figure}[t]
	\centering
	\begin{minipage}{0.9\textwidth}
		\centering
		\begin{tcolorbox}[colback=white!98!black, colframe=lightgray!45, boxrule=0.25pt, arc=2pt, left=4pt, right=4pt, top=4pt, bottom=2pt]
			\begin{minted}[fontsize=\scriptsize, breaklines, autogobble]{cpp}
void qemu_ram_free(struct uc_struct *uc, RAMBlock *block)
{
    if (!block) {
        return;
    }

    QLIST_REMOVE(block, next);
    uc->ram_list.mru_block = NULL;
    reclaim_ramblock(uc, block);
}
\end{minted}
		\end{tcolorbox}
	\end{minipage}
	\adjustbox{max width=\textwidth, keepaspectratio, trim=0 0 15pt 0}{
		\input{figures/grapht5/vdg_new}
	}
	\caption{\ac{vdg} of a C function containing a null pointer dereference (\CWE{476}) vulnerability~\cite{nationalvulnerabilitydatabase:2022:cve202229694}. The source excerpt is shown above the graph. Edge colors distinguish control-flow, control-dependence, and data-dependence edges. Data-dependence edges are labeled with the variable that carries the dependence (\texttt{block} or \texttt{uc}).}
	\Description{The qemu_ram_free function and its source-line graph. Directed control-flow, control-dependence, and data-dependence edges connect the statements around the null-pointer vulnerability.}
	\label{fig:impl:vdg}
\end{figure}
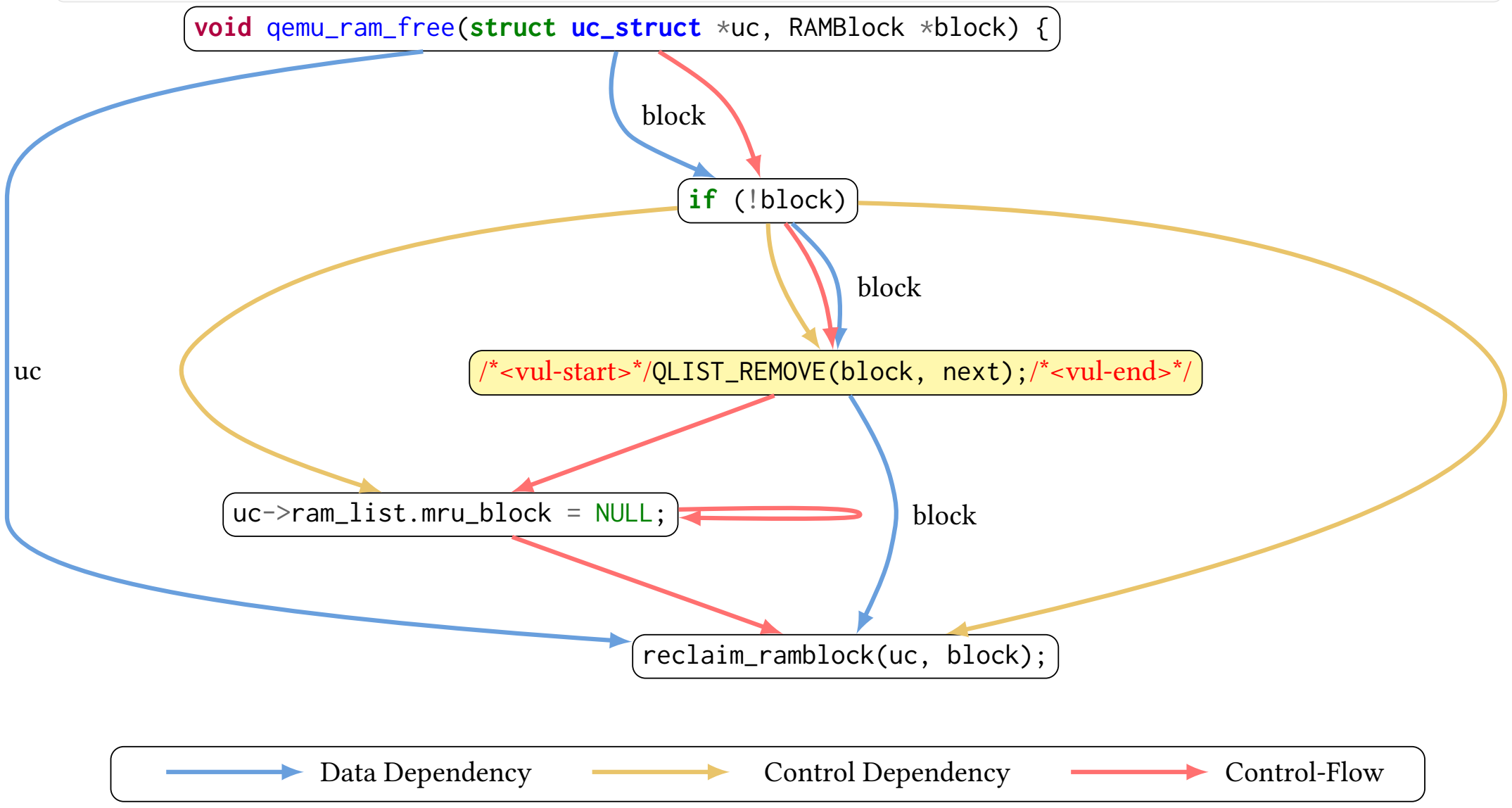

\ac{vdg} construction assumes perfect localization: each vulnerable hunk is marked with \texttt{<vul-start>} and \texttt{<vul-end>}. If a marked range contains no \ac{cpg} node, we extend it to the nearest lines before and after it that do. Slicing therefore starts from program constructs rather than from comments, blank lines, or marker-only lines.

We parse the marked function with \Joern~\cite{yamaguchi:sp2014:modeling} to build a \ac{cpg}. A \ac{cpg} represents code as a single graph: nodes are program elements, directed labeled edges are relations between them, and node attributes store element metadata. Because \Joern provides frontends for several languages, the same construction applies to every language it supports; we use its C/C++ and Java frontends.

Next, we merge all \ac{cpg} nodes on the same source line into one \ac{vdg} node, so each \ac{vdg} node corresponds to exactly one line. We keep a subset of the \ac{cpg} relations and project them onto these line nodes. Control-flow, control-dependence, and data-dependence edges form the backbone. We additionally keep \texttt{REF} edges for declaration-use relations relevant to the vulnerable region, and \texttt{AST} edges where they are needed to connect declarations to the surrounding code. The resulting \ac{vdg} can contain five edge types:
\begin{itemize}
	\item Abstract syntax tree (\texttt{AST}) edges, which connect retained declarations to the surrounding code structure.
	\item Control-flow (\texttt{CFG}) edges, which capture possible execution order.
	\item Control-dependence (\texttt{CDG}) edges, which capture that the execution of one line depends on a predicate in another line.
	\item Data-dependence (\texttt{DDG}, derived from \texttt{REACHING\_DEF}) edges, which capture that a definition reaches a later use of the same variable.
	\item Reference (\texttt{REF}) edges, which capture declaration-use relations for identifiers relevant to the vulnerable region.
\end{itemize}
To avoid duplicate links, we drop an \texttt{AST} or \texttt{REF} edge if a \texttt{CFG}, \texttt{CDG}, or \texttt{DDG} edge already connects the same ordered pair of lines. Each retained edge also stores the character ranges of its source and destination within the two lines. Nodes thus remain whole lines, while edges identify the sub-line spans involved in the dependency.

Finally, we perform \textbf{vulnerability-centric slicing}. The lines in the marked hunks seed a forward and a backward traversal over all retained edge types, including \texttt{REF} and \texttt{AST}. Forward slicing collects the code that depends on the vulnerable lines; backward slicing collects the code they depend on. We keep all nodes reached in either direction and the edges between them.
Figure~\ref{fig:impl:vdg} shows the \ac{vdg} of a vulnerable function from the Unicorn Engine, an open-source CPU emulator~\cite{thincasttechnologiesgmbh:2020:fixed}.

The result is a \textbf{source-aligned} graph centered on the vulnerability: each node is one line of code, and each edge records the source spans involved in the relation. \GraphTFive uses the spans to select tokens when truncating long lines and uses the line-level edges to restrict attention between lines (Section~\ref{sec:impl:statement-level-edge-projection}). The next subsection describes how \GraphTFive encodes these nodes and edges.

%% file: figures/grapht5/vdg_new.tex
\newcommand{\customlegend}[4]{%
    \node[anchor=center] at (#1, #2) {
        \begin{tikzpicture}[baseline=(current bounding box.center), scale=#3, transform shape]
            \node[draw, rounded corners, inner sep=3pt, align=center] (legendbox) {
                \hspace*{1.5em}%
                \begin{tabular}{@{}*{10}{c@{\hspace{0.5em}}r}@{}}
                    #4
                \end{tabular}
            };
        \end{tikzpicture}
    };
}

\newcommand{\legenditem}[2]{%
    \tikz[baseline]{\draw[#1, ->, line width=1.35pt] (0,0.1) -- (1.5,0.1);} & #2%
}

\newcommand{\legendsep}{\hspace{2em}}

\begin{tikzpicture}[>=latex,line join=bevel,yscale=0.6 ]
    \node (5) at (256.25bp,164.0bp) [draw,rectangle,rounded corners,fill=yellow!40] {\textcolor{red}{/*<vul-start>*/}\mintinline{cpp}|QLIST_REMOVE(block, next);|\textcolor{red}{/*<vul-end>*/}};
    \node (6) at (137.25bp,91.0bp) [draw,rectangle,rounded corners] {\mintinline{cpp}|uc->ram_list.mru_block = NULL;|};
    \node (7) at (259.25bp,18.0bp) [draw,rectangle,rounded corners] {\mintinline{cpp}|reclaim_ramblock(uc, block);|};
    \node (1) at (190.25bp,341.0bp) [draw,rectangle,rounded corners] {\mintinline{cpp}|void qemu_ram_free(struct uc_struct *uc, RAMBlock *block) {|};
    \node (2) at (235.25bp,252.5bp) [draw,rectangle,rounded corners] {\mintinline{cpp}|if (!block)|};
    \draw [pastelred,->,line width=1.35pt] (5) ..controls (211.72bp,136.43bp) and (192.59bp,125.02bp)  .. (6);
    \draw [pastelblue,->,line width=1.35pt] (5) ..controls (266.96bp,135.36bp) and (271.31bp,121.63bp)  .. (273.25bp,109.0bp) .. controls (275.68bp,93.186bp) and (275.27bp,88.871bp)  .. (273.25bp,73.0bp) .. controls (272.17bp,64.511bp) and (270.16bp,55.449bp)  .. (7);
    \definecolor{strokecol}{rgb}{0.0,0.0,0.0}
    \pgfsetstrokecolor{strokecol}
    \draw (289.92bp,91.0bp) node {block};
    \draw [pastelred,->,line width=1.35pt] (6) ..controls (256.89bp,95.462bp) and (263.75bp,93.602bp)  .. (263.75bp,91.0bp) .. controls (263.75bp,89.496bp) and (261.46bp,88.239bp)  .. (6);
    \draw [pastelred,->,line width=1.35pt] (6) ..controls (182.9bp,63.434bp) and (202.52bp,52.017bp)  .. (7);
    \draw [pastelblue,->,line width=1.35pt] (1) ..controls (28.236bp,310.13bp) and (0.25bp,289.11bp)  .. (0.25bp,253.5bp) .. controls (0.25bp,253.5bp) and (0.25bp,253.5bp)  .. (0.25bp,90.0bp) .. controls (0.25bp,56.697bp) and (80.904bp,38.453bp)  .. (7);
    \draw (6.625bp,164.0bp) node {uc};
    \draw [pastelblue,->,line width=1.35pt] (1) ..controls (185.92bp,312.44bp) and (186.02bp,299.08bp)  .. (191.25bp,288.5bp) .. controls (192.99bp,284.98bp) and (195.27bp,281.7bp)  .. (2);
    \draw (206.25bp,296.75bp) node {block};
    \draw [pastelred,->,line width=1.35pt] (1) ..controls (213.13bp,317.39bp) and (217.86bp,311.29bp)  .. (221.25bp,305.0bp) .. controls (225.05bp,297.96bp) and (227.88bp,289.81bp)  .. (2);
    \draw [pastelblue,->,line width=1.35pt] (2) ..controls (250.6bp,228.7bp) and (253.57bp,222.57bp)  .. (255.25bp,216.5bp) .. controls (257.22bp,209.35bp) and (258.02bp,201.42bp)  .. (5);
    \draw (272.82bp,208.25bp) node {block};
    \draw [pastelyellow,->,line width=1.35pt] (2) ..controls (235.46bp,222.64bp) and (239.04bp,206.73bp)  .. (5);
    \draw [pastelred,->,line width=1.35pt] (2) ..controls (249.01bp,222.76bp) and (253.12bp,207.05bp)  .. (5);
    \draw [pastelyellow,->,line width=1.35pt] (2) ..controls (146.48bp,240.45bp) and (88.617bp,223.79bp)  .. (60.25bp,182.0bp) .. controls (51.264bp,168.76bp) and (52.867bp,160.19bp)  .. (60.25bp,146.0bp) .. controls (66.666bp,133.66bp) and (77.333bp,123.56bp)  .. (6);
    \draw [pastelyellow,->,line width=1.35pt] (2) ..controls (333.79bp,248.52bp) and (413.54bp,235.04bp)  .. (452.25bp,182.0bp) .. controls (496.73bp,121.05bp) and (394.26bp,68.406bp)  .. (7);
    \customlegend{235.25bp}{-1.5}{1}{
        \legenditem{pastelblue}{Data Dependency}\legendsep
        \legenditem{pastelyellow}{Control Dependency}\legendsep
        \legenditem{pastelred}{Control-Flow}
    }%
\end{tikzpicture}

%% file: sections/grapht5.tex
\begin{figure}[t]
	\centering
	\includegraphics[width=\textwidth]{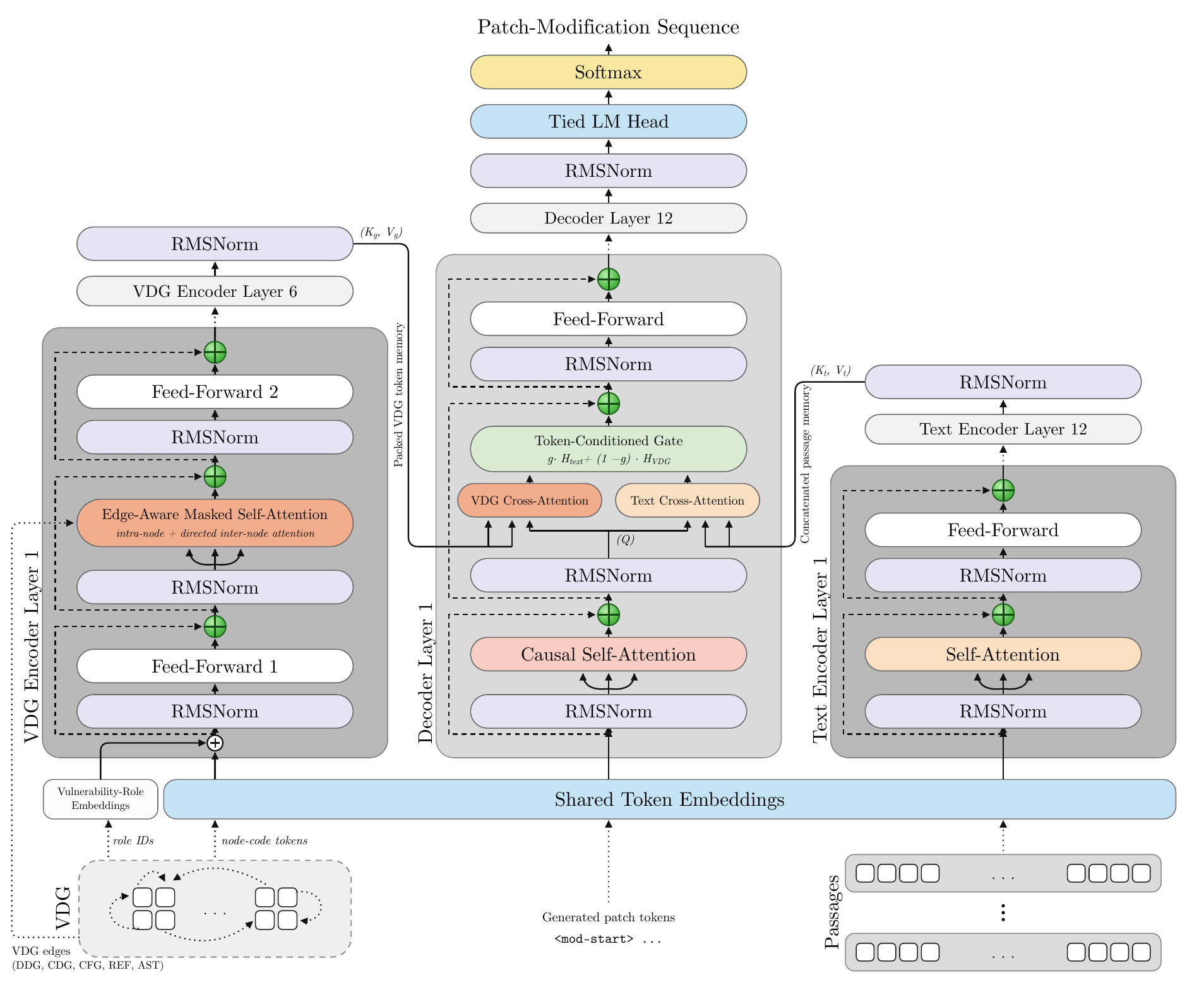}
	\caption{Architecture of \GraphTFive. The text encoder processes source-code passages independently, while the \ac{vdg} encoder combines attention within nodes with edge-constrained attention between nodes. Each decoder layer attends separately to both memories and combines them through a token-dependent gate.}
	\Description{A CodeT5 text encoder and a six-layer edge-aware graph encoder produce separate memories. Each decoder layer combines text and graph cross-attention outputs through a token-conditioned gate.}
	\label{fig:impl:grapht5-network-architecture}
\end{figure}

\GraphTFive uses two encoders and one autoregressive decoder.
The text encoder, initialized from \CodeTFive, encodes passages of the vulnerable function independently and preserves their sequential context.
The \ac{vdg} encoder contextualizes tokens within each line node and across directed \ac{vdg} edges.
Each decoder layer attends separately to the text and graph memories and combines the two results through a learned, token-dependent gate.
Figure~\ref{fig:impl:grapht5-network-architecture} summarizes the two encoders and the \ac{dca} decoder.

\subsubsection{\ac{vdg} Encoder}

The \ac{vdg} encoder turns the line-level graph into contextual token representations for the decoder.
It preserves the node structure: tokens can always attend to tokens in their own line, and attention between lines is allowed only along directed \ac{vdg} edges.
The encoder has six layers, each consisting of a feed-forward sublayer, an edge-aware masked self-attention sublayer, and a second feed-forward sublayer.

\paragraph{\textbf{Node Representations}}
As described in Section~\ref{sec:the-vulnerability-dependence-graph}, each \ac{vdg} node corresponds to one source line.
We tokenize the line with the \CodeTFive tokenizer and embed its tokens with the embedding table shared across the model.
Each node also receives a binary role, vulnerable or context, derived from the marked ranges.
The learned role embedding is added to every token embedding of the node before the first encoder layer.
This marks vulnerable nodes without inserting marker tokens into each node.

Each node holds at most \num{40} tokens.
For longer lines, the data loader first keeps one representative token from each vulnerability-marker span and each edge-endpoint span.
It then keeps the remaining tokens of these spans and fills any free positions with nearby tokens.
The spans thus only guide truncation; they do not define token-level edges.

\paragraph{\textbf{Line-Level Edge Projection}}
\label{sec:impl:statement-level-edge-projection}
The data loader maps the source and destination lines of each retained \ac{vdg} edge to positions in the selected node sequence.
It removes self-loops and duplicate edges with the same source, destination, and type.
Each edge stores its source index, destination index, type, and the signed line offset between its endpoints.

For a directed edge \(u \rightarrow v\), every valid token in node \(v\) may attend to every valid token in node \(u\).
Attention thus follows the direction of the edge, and both lines remain fully visible.

\begin{figure}[t]
	\centering

	\adjustbox{max width=\textwidth, keepaspectratio, trim=0 0 15pt 0}{
		\includegraphics[]{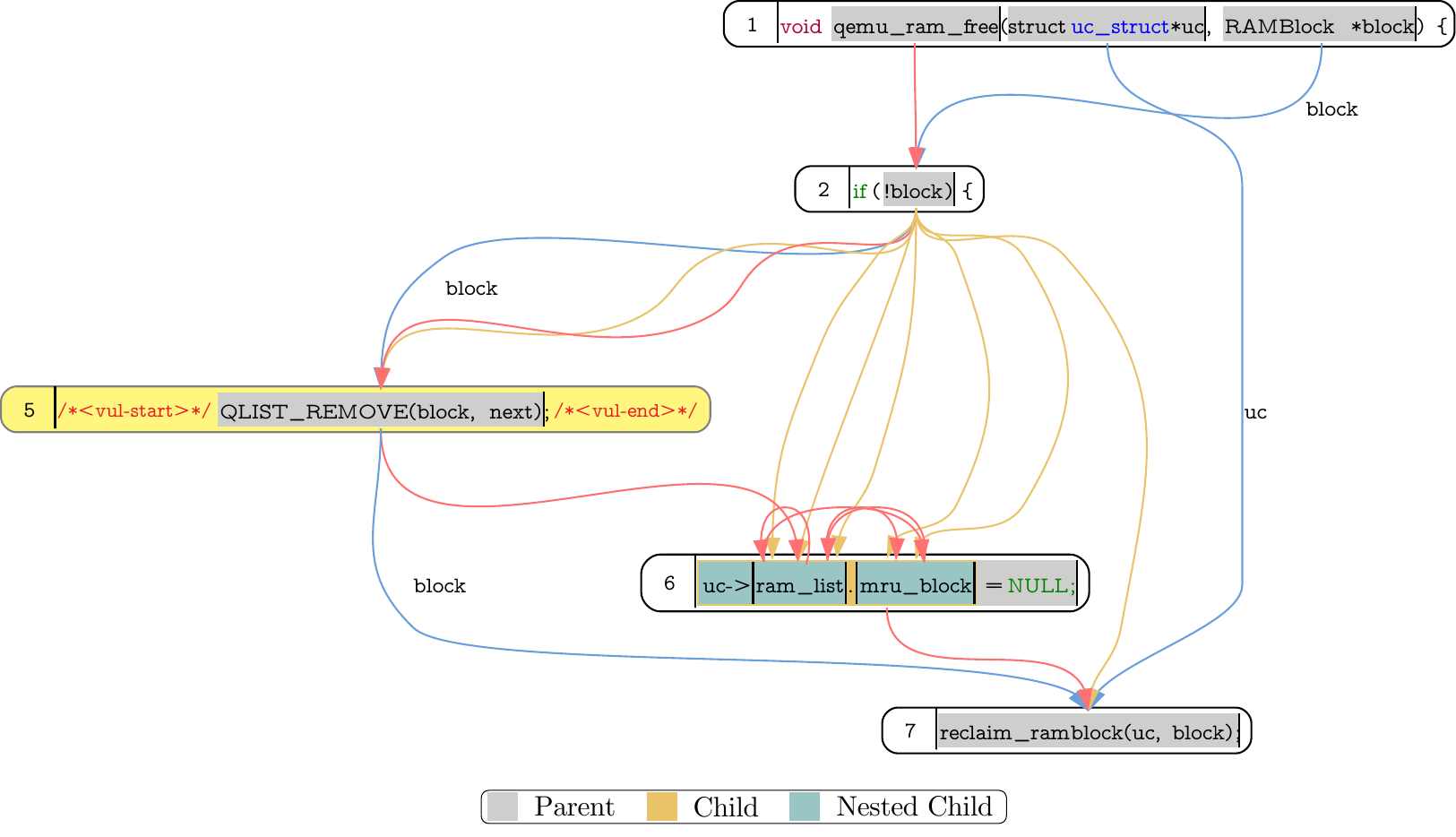}
	}

	\caption{Source-span annotations in a \ac{vdg} for a C function containing a null pointer dereference (\CWE{476}). The colored subsequences mark the character spans stored for the endpoints of each relation. \GraphTFive uses these spans to select tokens when a line exceeds the node-token budget.}
	\Description{Source lines form graph nodes. Colored token spans within the lines identify the source and destination annotations of directed graph relations.}
	\label{fig:impl:vdg-source-spans}
\end{figure}

\paragraph{\textbf{Edge Type and Distance}}\label{subsubsec:impl:relative-position-bias}
Each line-level edge contributes two learned scalar biases per attention head.
The type bias distinguishes the five edge types: \texttt{AST}, \texttt{CFG}, \texttt{CDG}, \texttt{DDG}, and \texttt{REF}.
The distance bias is indexed by a logarithmic bucket of the signed line offset and therefore encodes both direction and approximate distance.
With \num{12} attention heads, the model learns an edge-type table in \(\mathbb{R}^{5 \times 12}\) and a distance table in \(\mathbb{R}^{32 \times 12}\).
Both biases are added to the corresponding attention logits.

\paragraph{\textbf{Unified Local and Edge-Aware Attention}}\label{sec:impl:unified-edge-aware-attention}
For attention, the node-token tensor is flattened, and the fixed node layout determines the attention mask.
Tokens within the same node attend to each other with \TFive's relative-position bias.
Across nodes, a token in a destination line may attend only to tokens in source lines connected to it by a \ac{vdg} edge; for these pairs, the edge-type and distance biases replace the positional bias.
A single masked self-attention operation therefore covers both attention within lines and message passing between lines.

Each encoder layer applies Feed-Forward~1, edge-aware masked self-attention, and Feed-Forward~2, each with pre-normalization and a residual connection.
Feed-Forward~1 transforms token and role information before any information crosses line boundaries; Feed-Forward~2 transforms the contextualized representations.
A final RMSNorm after the sixth layer produces the graph memory for the decoder.

\paragraph{\textbf{Implementation Details}}
Token IDs are embedded with the \CodeTFive embedding layer, which is shared by the text encoder, \ac{vdg} encoder, decoder, and output head.
A graph with \(N\) selected nodes yields features of shape \([N,L_{seq},d]\), where \(L_{seq}=40\) and \(d=768\).
The graph contains at most \num{40} nodes.
For larger graphs, the model keeps all vulnerable nodes and fills the remaining positions in order of increasing graph distance from them.
Within each distance, it prefers highly connected nodes, gives data-dependence edges additional weight, and breaks ties by source-line proximity.
The only exception to the limit is a vulnerable region with more than \num{40} lines, in which case all vulnerable nodes are kept.

We set \(L_{seq}=40\) based on the node-length distribution of \BigVul + \CVEfixes under \CodeTFive tokenization: median \(14\), mean \(17.25\), and 90th percentile \(33\) tokens.
The limit therefore truncates only unusually long lines.
For attention, the node features are flattened to \([(N \times L_{seq}) \times d]\).
The fixed layout determines node membership, and node masks mark valid and padding positions.
After encoding, the valid token representations are packed contiguously and passed to the decoder with the corresponding graph-memory mask.

The six-layer \ac{vdg} encoder is initialized from the twelve layers of the pretrained \CodeTFive encoder. For each consecutive pair of layers, Feed-Forward~1 is initialized from the first layer, and the self-attention and Feed-Forward~2 sublayers from the second. The edge biases and role embeddings are new parameters trained from scratch.

\subsubsection{Text Context and Repair Instructions}

The text encoder gives \GraphTFive the full function context, including tokens that are absent from the graph but needed to interpret or locate an edit.
The model outputs repairs in the modification grammar of Chen et al.~\cite{chen:tse2023:neural} rather than as line-based diffs.
\VulMaster~\cite{zhou:icse2024:out} uses the same target representation, so both models are compared on identical targets in Section~\ref{sec:evaluation}.

\begin{figure}[t]
	\centering
	\input{figures/background/modification/patch_mod_diff}
	\caption{How the modification grammar expresses line diffs as patch instructions. Edits are delimited by \texttt{<mod-start>} and, for deletions and replacements, \texttt{<mod-end>}.}
	\Description{Insertion, deletion, and replacement examples show how unchanged context tokens anchor modifications delimited by the mod-start and mod-end markers.}
	\label{fig:back:patch-modification-syntax}
\end{figure}

Instead of regenerating every modified line, even for changes of a few characters, the grammar encodes each edit with a fixed-size \textit{context window} around the changed span. This allows precise in-line edits and shortens the target sequences, which matters because \ac{s2s} models typically degrade as output length grows.

Each edit has one of three types and is delimited by \texttt{<mod-start>} and, for DELETE and REPLACE, \texttt{<mod-end>}. Each instruction includes unchanged context tokens that anchor the edit location. The placeholder \texttt{new} denotes the generated repair tokens:

\begin{enumerate}
	\item \textbf{ADD}:
	      \texttt{<mod-start>} \textit{pre-context} \textit{new}

	\item \textbf{DELETE}:
	      \texttt{<mod-start>} \textit{pre-context} \texttt{<mod-end>} \textit{post-context}

	\item \textbf{REPLACE}:
	      \texttt{<mod-start>} \textit{pre-context} \textit{new} \texttt{<mod-end>} \textit{post-context}
\end{enumerate}

The context tokens are not modified; they only locate the edit. With this grammar, \GraphTFive generates a sequence of edits that changes only the vulnerable regions, including repairs that span multiple hunks.

Because instructions are anchored by context tokens rather than line numbers, tokens unrelated to the vulnerability can still be needed to place an edit.
We therefore adopt \acf{fid}~\cite{izacard:eacl2021:leveraging}, which avoids the encoder's input-length limit by encoding multiple passages independently and concatenating their hidden states for decoding. We split the input function into passages, encode each passage with the pretrained \CodeTFive encoder~\cite{wang:emnlp2021:codet5}, and pass all encoded passages to the decoder.
This keeps the full context needed to place instructions and preserves the pretrained behavior of the text encoder.

\subsubsection{Modality Fusion in the Decoder}\label{sec:impl:modality-fusion-in-the-decoder}

Following the \ac{fid} principle of fusing separately encoded contexts in the decoder~\cite{izacard:eacl2021:leveraging}, we replace the cross-attention in \TFive's decoder with a dual cross-attention module.
The text encoder produces hidden states \(H^{\text{text}}=[h_1,\ldots,h_m]\), where \(m\) is the number of text tokens across all passages. The graph encoder produces hidden states \(H^{\text{graph}}=[g_1,\ldots,g_n]\), where \(n\) is the number of valid node tokens after packing.
Each decoder layer performs two parallel cross-attention operations, one over \(H^{\text{text}}\) and one over \(H^{\text{graph}}\), and combines their outputs through a token-dependent learned gate:

\begin{equation}
	\begin{aligned}
		g_t^\ell         & = \sigma\!\left(
		\alpha^\ell + (\mathbf{w}^\ell)^\top
		\left[
			\operatorname{RMSNorm}(h_t^\ell);
			c_{t,\text{text}}^\ell;
			c_{t,\text{graph}}^\ell
			\right]
		\right),                                                       \\
		\tilde{h}_t^\ell & = h_t^\ell + \operatorname{Dropout}\!\left(
		g_t^\ell c_{t,\text{text}}^\ell +
		(1-g_t^\ell)c_{t,\text{graph}}^\ell
		\right).
	\end{aligned}
	\label{seq:impl:eq:cross-attn}
\end{equation}

where \(c_{t,\text{text}}^\ell\) and \(c_{t,\text{graph}}^\ell\) are the two cross-attention context vectors, \(\alpha^\ell\) is a learned per-layer baseline, and \(\mathbf{w}^\ell\) maps the normalized decoder state and both context vectors to a token-specific adjustment. The text weight \(g_t^\ell\) can therefore vary across samples and decoder positions, and the graph branch receives \(1-g_t^\ell\).
Before fusion, the text memory reflects \TFive's relative-position modeling, and the graph memory reflects the edge-constrained mask and the relation biases.
The two cross-attention operations have separate parameters and masks and add no relative-position biases of their own.
The graph cross-attention projections are initialized from the text cross-attention projections, and the gate is initialized to weight both modalities equally.

%% file: figures/background/modification/patch_mod_diff.tex
\usemintedstyle{tango}

\definecolor{lightgray}{RGB}{248,248,248}
\definecolor{mediumgray}{HTML}{F5F5F7}
\definecolor{bordercolor}{RGB}{204,204,204}
\definecolor{arrowcolor}{HTML}{B4B4B8}
\definecolor{deletecolor}{RGB}{255,238,238}
\definecolor{replacecolor}{RGB}{255,243,224}
\definecolor{addcolor}{RGB}{238,255,238}

\definecolor{diffadd}{HTML}{E6FFEC}
\definecolor{diffdelete}{HTML}{FFEBE9}

\newtcolorbox{diffbox}[2][]{%
    enhanced jigsaw,
    halign=center, valign=center,
    title=#2,
    colback=lightgray,
    colframe=bordercolor,
    fonttitle=\scshape,
    coltitle=black,
    colbacktitle=mediumgray,
    width=\textwidth,
    boxsep=2pt,
    right=0pt, top=0pt, bottom=0pt,
    boxrule=0.5pt,
    #1
}

\begin{tikzpicture}[
        font=\sffamily,
        scale=0.85,
        transform shape,
        box/.style={anchor=center, inner sep=0pt, outer sep=0pt, blur shadow={shadow opacity=20, shadow blur extra rounding, shadow blur radius=1ex, shadow blur steps=6}},
        header/.style={rectangle, rounded corners=5pt, draw=bordercolor, line width=1pt, fill=mediumgray},
        tablecell/.style={rectangle, draw=bordercolor, line width=0.5pt, fill=lightgray},
        tableheader/.style={rectangle, draw=bordercolor, line width=0.5pt, fill=mediumgray},
        deleterow/.style={rectangle, draw=bordercolor, line width=0.5pt, fill=deletecolor},
        replacerow/.style={rectangle, draw=bordercolor, line width=0.5pt, fill=pastelyellow!30!white},
        addrow/.style={rectangle, draw=bordercolor, line width=0.5pt, fill=pastelgreen!30!white},
        code/.style={font=\ttfamily\small}
    ]

    \node[box] (box1) at (0.0,0.1) {\input{figures/background/modification/vulnerable_code}};

    \draw[->, line width=.75pt, arrowcolor, blur shadow={shadow opacity=8, shadow blur extra rounding, shadow blur radius=0.55ex}] (0,-1.55) -- (0,-2);

    \node[
        blur shadow={shadow opacity=20, shadow blur extra rounding, shadow blur radius=1ex, shadow blur steps=6},
        inner sep=0pt
    ] at (0,-3.5) {
        \begin{tikzpicture}[font=\ttfamily\small]
            \draw[tableheader]
            (-6.75,-1.75) -- (-4.125,-1.75)
            -- (-4.125,-2.25)
            -- (-6.75,-2.25) {[rounded corners=6pt] -- cycle};
            \draw[tableheader] (-4.125,-1.75) rectangle (-3.1875,-2.25);
            \draw[tableheader] (-3.1875,-1.75) rectangle (-0.375,-2.25);
            \draw[tableheader] (-0.375,-1.75) rectangle (3,-2.25);
            \draw[tableheader]
            (3,-1.75) {[rounded corners=6pt] -- (6.75,-1.75)}
            {[sharp corners] -- (6.75,-2.25)}
            -- (3,-2.25)
            -- cycle;
            \node[font=\scshape,scale=0.75] at (-5.4375,-2) {OPERATION};
            \node[font=\scshape,scale=0.75] at (-3.65625,-2) {\#};
            \node[font=\scshape,scale=0.75] at (-1.78125,-2) {Pre-Context};
            \node[font=\scshape,scale=0.75] at (1.3125,-2) {Modification};
            \node[font=\scshape,scale=0.75] at (4.875,-2) {Post-Context};

            \draw[deleterow] (-6.75,-2.25) rectangle (-4.125,-3);
            \draw[tablecell] (-4.125,-2.25) rectangle (-3.1875,-3);
            \draw[tablecell] (-3.1875,-2.25) rectangle (-0.375,-3);
            \draw[tablecell] (-0.375,-2.25) rectangle (3,-3);
            \draw[tablecell] (3,-2.25) rectangle (6.75,-3);
            \node[font=\rmfamily,scale=0.75] at (-5.4375,-2.625) {DELETE};
            \node[scale=0.75] at (-3.65625,-2.625) {1};
            \node[code,scale=0.75] at (-1.78125,-2.625) {getSalt ( )};
            \node[code,scale=0.75] at (1.3125,-2.625) {$\varnothing$};
            \node[code,scale=0.75] at (4.875,-2.625) {\{ byte [};

            \draw[replacerow] (-6.75,-3) rectangle (-4.125,-3.75);
            \draw[tablecell] (-4.125,-3) rectangle (-3.1875,-3.75);
            \draw[tablecell] (-3.1875,-3) rectangle (-0.375,-3.75);
            \draw[tablecell] (-0.375,-3) rectangle (3,-3.75);
            \draw[tablecell] (3,-3) rectangle (6.75,-3.75);
            \node[font=\rmfamily,scale=0.75] at (-5.4375,-3.375) {REPLACE};
            \node[scale=0.75] at (-3.65625,-3.375) {2};
            \node[code,scale=0.75] at (-1.78125,-3.375) {] salt =};
            \node[code,scale=0.75] at (1.3125,-3.375) {new byte [};
            \node[code,scale=0.75] at (4.875,-3.375) {this . saltLength};

            \draw[addrow]
            (-6.75,-3.75) -- (-4.125,-3.75)
            -- (-4.125,-4.5)
            {[rounded corners=6pt] -- (-6.75,-4.5)}
            -- cycle;
            \draw[tablecell] (-4.125,-3.75) rectangle (-3.1875,-4.5);
            \draw[tablecell] (-3.1875,-3.75) rectangle (-0.375,-4.5);
            \draw[tablecell] (-0.375,-3.75) rectangle (3,-4.5);
            \draw[tablecell]
            (3,-3.75) -- (6.75,-3.75)
            {[rounded corners=6pt] -- (6.75,-4.5)}
            {[sharp corners] -- (3,-4.5)}
            -- cycle;
            \node[font=\rmfamily,scale=0.75] at (-5.4375,-4.125) {ADD};
            \node[scale=0.75] at (-3.65625,-4.125) {3};
            \node[code,scale=0.75] at (-1.78125,-4.125) {this . saltLength};
            \node[code,scale=0.75] at (1.3125,-4.125) {]; sr.nextBytes(salt)};
            \node[code,scale=0.75] at (4.875,-4.125) {$\varnothing$};%
        \end{tikzpicture}
    };

    \draw[->, line width=0.75pt, arrowcolor, blur shadow={shadow opacity=8, shadow blur extra rounding, shadow blur radius=0.55ex}] (0,-5) -- (0,-5.45);

    \node[box] (box2) at (0,-7.1) {\input{figures/background/modification/patched_code}};
\end{tikzpicture}

%% file: sections/evaluation.tex
This section evaluates \GraphTFive in two settings. A static evaluation with similarity metrics uses three large C/C++ datasets: \BigVul + \CVEfixes~\cite{zhou:icse2024:out}, \ICVul~\cite{lu:2025:icvul}, and \PrimeVul~\cite{ding:icse2025:vulnerability}. A dynamic validation with exploits and tests uses the C/C++ benchmark \datasetname{Vul4C}~\cite{hu:2025:sok} and the Java benchmark \vulfourj~\cite{bui:msr2022:vul4j}.

\subsection{Evaluation Design}\label{sec:eval:research-questions}

We address the following research questions:

\begin{enumerate}[label=\textbf{RQ\textsubscript{\arabic*}}]
	\item \textbf{Static comparison.} How does \GraphTFive compare with state-of-the-art \ac{avr} approaches and closed \acp{llm} on C/C++ vulnerabilities under static similarity metrics?
	\item \textbf{Number of repair hunks.} How does repair performance of \GraphTFive vary with the number of repair hunks?
	\item \textbf{Performance across CWE types.} How does the repair performance of \GraphTFive on C/C++ vary across \ac{cwe} types?
	\item \textbf{Dynamic validation on C/C++.} How effectively does \GraphTFive repair reproducible C/C++ vulnerabilities when its patches are validated with exploits and regression tests?
	\item \textbf{Java adaptation.} After fine-tuning on Java, how effectively does \GraphTFive repair Java vulnerabilities when its patches are validated with \ac{pov} tests?
\end{enumerate}

\paragraph{\textbf{Datasets}}
The static evaluation uses three C/C++ datasets. The deduplicated \BigVul + \CVEfixes split~\cite{zhou:icse2024:out} contains \num{5800} function-level fixes from \num{1754} projects across \num{96} \ac{cwe} categories.
\ICVul contains \num{6276} vulnerable functions across \num{146} \ac{cwe} categories~\cite{lu:2025:icvul}. \PrimeVul contains \num{6968} vulnerable functions across \num{140} \ac{cwe} categories and is split chronologically to reduce temporal leakage~\cite{ding:icse2025:vulnerability}. These numbers refer to the source corpora before our preprocessing. After preprocessing and splitting, the C/C++ test set contains \num{2411} samples: \num{1496} from \BigVul + \CVEfixes, \num{487} from \ICVul, and \num{428} from \PrimeVul. Table~\ref{tab:eva:model-comparison}, Figures~\ref{fig:eva:hunk-comparison-best-models} and~\ref{fig:eva:comparison-best-models}, and Appendix Figure~\ref{app:fig:cwe-similarity-scores} use these test samples.

\paragraph{\textbf{Static Evaluation}}
\rqOne--\rqThree use three static similarity metrics. \EM measures exact agreement with the developer patch, \BLEUw measures token-level n-gram overlap~\cite{papineni:acl2002:bleu}, and \CodeBLEU combines lexical, syntactic, and data-flow similarity~\cite{ren:arxiv2020:codebleu}. Table~\ref{tab:eva:model-comparison} reports corpus-level scores. The structural components of \CodeBLEU are less informative on instructions than on complete programs (Section~\ref{sec:discussion_evaluation}).

A hunk is a contiguous block of changed lines in the developer patch. \rqTwo separates single-hunk from multi-hunk repairs and reports \EM per hunk count. We also report the hunk-weighted exact match \(\text{EM}_w = \frac{\sum_j m_j e_j}{\sum_j m_j}\), where \(m_j\) is the number of hunks in the target repair of sample \(j\) and \(e_j\) indicates an exact match. This metric gives more weight to repairs with more hunks.

All metrics are computed on modification-instruction sequences. \GraphTFive and \VulMaster generate these instructions directly. For the closed \acp{llm}, we derive instructions by diffing the returned function against the vulnerable input on the tokenized, \ac{ast}-normalized representation. Because the grammar operates on the \ac{ast}, this derivation ignores formatting: reformatting unmodified code produces no instructions. For \EM, we additionally accept a match at the full-function level for all four systems, after applying the generated instructions where applicable. This covers equivalent edits that are encoded differently from the reference, such as a REPLACE versus an equivalent DELETE followed by an ADD. The replication package contains the diffing and reconstruction code.

Static similarity metrics are the main large-scale evaluation oracle of prior learning-based \ac{avr} systems~\cite{harer:neurips2018:learning,zhou:emse2022:spvf,chi:tse2022:seqtrans,fu:fse2022:vulrepair,chen:tse2023:neural,fu:tosem2024:vision,zhou:icse2024:out,de-fitero-dominguez:eaai2024:enhanced,wen:2025:vulr2,wang:ase2026:spvr} (Table~\ref{tab:prestudy-avr}). Dynamic validation in source-level \ac{avr} is limited to smaller curated benchmarks or to settings with \ac{poc} exploits~\cite{bui:saner2026:vulterminator,gao:2021:tests,zhang:issta2022:program,kulsum:aiware2024:case,nong:uss2025:appatch,kim:uss2025:logs,yu:uss2025:patchagent,zhang:icse-seip2026:fixing,ye:arxiv2026:well}.
We therefore use similarity metrics for the large-scale C/C++ comparison and complement them with dynamic validation in \rqFour and \rqFive.

\paragraph{\textbf{Dynamic Validation}}
For \rqFour, we use \datasetname{Vul4C}, which contains \num{144} reproducible C/C++ vulnerabilities from \num{23} projects and \num{19} \ac{cwe} categories~\cite{hu:2025:sok}. Each vulnerability comes with a \ac{poc} exploit that triggers it in the original program, the developer patch, and, where available, regression tests. Hu et al. evaluate learning-based \ac{avr} models on the \num{81} single-file vulnerabilities that these models can process, with \num{50} candidates per vulnerability. We follow this setup: \GraphTFive generates \num{50} candidates for each of the \num{81} vulnerabilities. We compile each candidate and run the \ac{poc} exploit and the available regression tests on it. A candidate is plausible if it compiles, prevents the exploit from triggering the vulnerability, and passes the regression tests. We report the compilation rate, the share of the \num{81} vulnerabilities with at least one compilable candidate, and pass@50~\cite{chen:2021:evaluating}, the share with at least one plausible candidate.

For \rqFive, we use the \num{35} single-hunk vulnerabilities of \vulfourj. We restrict the comparison to these vulnerabilities because prior Java \ac{avr} results are reported only for them~\cite{wu:issta2023:how,zhou:icse2024:out,bui:saner2026:vulterminator}. Wu et al.~\cite{wu:issta2023:how} selected them as the reproducible single-hunk vulnerabilities of \vulfourj, because the repair models they evaluate fix only single-hunk bugs. \vulfourj provides \ac{pov} tests for each vulnerability. A patch that passes these tests is plausible with respect to the available test suite. We report pass@10~\cite{chen:2021:evaluating}: the share of the \num{35} vulnerabilities for which at least one of ten candidates compiles and passes all \ac{pov} tests. Ten candidates is the budget of prior \vulfourj evaluations~\cite{wu:issta2023:how,bui:saner2026:vulterminator}; Wu et al. motivate it with the observation that developers examine at most ten candidate patches.

\paragraph{\textbf{Baselines}}
Across all research questions, we compare \GraphTFive with other \ac{avr} approaches.
For \rqOne--\rqThree, we compare with \VulMaster~\cite{zhou:icse2024:out}, the state of the art for C/C++ vulnerability repair. In a comprehensive evaluation, Wang et al.~\cite{wang:tosem2026:evaluation} found that \VulMaster outperforms other recently published \ac{avr} approaches, including \VRepair, \VulRepair, and \VQM; we therefore do not re-run these models. We retrain \VulMaster on the same splits as \GraphTFive, using the replication package of Zhou et al.~\cite{zhou:icse2024:out} with its default hyperparameters. \rqOne additionally includes two proprietary \acp{llm}, \GPTFourO and \GeminiTwoFivePro, to extend the comparison beyond task-specific models.
For \rqFour, we compare with the four learning-based \ac{avr} models that Hu et al.~\cite{hu:2025:sok} evaluate on \datasetname{Vul4C}: \VRepair, \VulRepair, \VQM, and \VulMaster. Their results are derived from the outcomes reported by Hu et al.
For \rqFive, we compare with the \vulfourj results reported by Wu et al.~\cite{wu:issta2023:how}, Bui et al.~\cite{bui:saner2026:vulterminator}, and Zhou et al.~\cite{zhou:icse2024:out}.

We evaluated both closed models zero-shot and deterministically on the \BigVul + \CVEfixes, \ICVul, and \PrimeVul test sets, using the snapshots \modelid{gpt-4o-2024-08-06} and \modelid{gemini-2.5-pro-2025-06-17}. For each sample, the prompt contained the \ac{cwe} identifier and the full vulnerable function, with the vulnerable lines marked by \texttt{/*<vul-start>*/} and \texttt{/*<vul-end>*/}. The model was instructed to return exactly one repaired function. We generated one candidate per sample with temperature \num{0.0} and top\_p \num{1.0} and evaluated the returned functions with the same metrics as the other models. %
Appendix~\ref{app:llm-prompt} shows the prompt template. Section~\ref{sec:validity} discusses possible benchmark exposure of the closed \acp{llm}.

\subsection{Training Setup}\label{sec:eval:training-setup}

We pretrain \GraphTFive for \num{8} epochs on the C/C++ bug-fix corpus by Chen et al.~\cite{chen:tse2023:neural}. We use AdamW~\cite{loshchilov:iclr2019:decoupled} with weight decay \num{0.1}. Each GPU processes two samples per batch. We accumulate gradients over \num{32} batches on each of two GPUs, giving an effective batch size of \num{128}. During pretraining, the text encoder processes up to three passages of \num{512} tokens. Pretraining takes approximately \num{50} hours. Pretraining and fine-tuning are supervised: the model learns to generate the target modification instructions by minimizing the token-level cross-entropy loss with teacher forcing.

\begin{table}[H]
	\centering
	\caption{Java fine-tuning data, drawn from the corpus of Bui et al.~\cite{bui:saner2026:vulterminator}.}
	\small
	\begin{tabular}{@{}lrrr@{}}
		\toprule
		                            & Training   & Validation & Combined   \\
		\midrule
		Repair examples             & \num{5295} & \num{589}  & \num{5884} \\
		GitHub repositories         & \num{1216} & \num{365}  & \num{1277} \\
		Unique commit hashes        & \num{4689} & \num{573}  & \num{5149} \\
		Unique source--target pairs & \num{4422} & \num{565}  & \num{4838} \\
		\bottomrule
	\end{tabular}
	\label{tab:eva:java-data}
\end{table}

The graph encoder and the graph-specific decoder components use a peak learning rate of \(4 \times 10^{-4}\), with linear warmup over the first \num{625} optimizer updates and linear decay afterwards. The text encoder, the text-path decoder parameters, and the shared embeddings stay frozen for the first \num{625} updates. Their learning rate then warms up to \(3 \times 10^{-5}\) over the next \num{625} updates and decays linearly.

For each C/C++ dataset, we fine-tune the pretrained model for \num{20} epochs with an effective batch size of \num{64}. Both parameter groups use a peak learning rate of \(4 \times 10^{-4}\) with linear warmup and decay. The text encoder processes up to \num{15} passages of \num{512} tokens, that is, up to \num{7680} tokens of context. We select the checkpoint with the highest mean of validation \EM and \CodeBLEU.

For \rqFive, we start from the C/C++-pretrained checkpoint and fine-tune it with the same hyperparameters on Java vulnerability fixes from the corpus of security-related GitHub commits that Bui et al.\ released with \VulTerminator~\cite{bui:saner2026:vulterminator}. Bui et al.\ collected this corpus from popular Java repositories: they matched commit messages against security-related keywords and kept commits whose messages start with a fixing verb, such as \emph{fix}, \emph{prevent}, or \emph{validate}. They also removed all commits of their test datasets, which include \vulfourjplus and therefore the \num{35} \vulfourj vulnerabilities of \rqFive. We matched the repository and commit of every fine-tuning example to this released corpus, so no example comes from a fix commit of a \vulfourj test vulnerability. Table~\ref{tab:eva:java-data} summarizes the fine-tuning data. The largest contributors are Apache OFBiz (\num{77} examples), Liferay Portal (\num{75}), Apache Tomcat (\num{68}), ResurrectionRemix Settings (\num{64}), and Jenkins and Spring Security (\num{63} each).

\begin{table}[H]
	\centering
	\caption{Comparison of \GraphTFive, \VulMaster, \GPTFourO, and \GeminiTwoFivePro on C/C++ vulnerabilities. Scores are corpus-level. Paired tests on the combined test set compare per-sample outcomes: McNemar's exact test for \EM and two-sided Wilcoxon signed-rank tests for \BLEUw and \CodeBLEU. Largest \(p\)-value across metrics: vs.\ \VulMaster, \(p \le 1.11 \times 10^{-3}\); vs.\ \GPTFourO, \(p \le 4.4 \times 10^{-51}\); vs.\ \GeminiTwoFivePro, \(p \le 5.6 \times 10^{-36}\).}
	\footnotesize
	\setlength{\tabcolsep}{3.5pt}
	\begin{tabular}{@{}llccccccccc@{}}
		\toprule
		\multirow{2}{*}{Type}                                                                                                                                                                                                                                                                                           & \multirow{2}{*}{Model}                & \multicolumn{3}{c}{\makecell[c]{\BigVul + \CVEfixes\\\scriptsize test \(n=\num{1496}\)}} & \multicolumn{3}{c}{\makecell[c]{\ICVul\\\scriptsize test \(n=\num{487}\)}} & \multicolumn{3}{c}{\makecell[c]{\PrimeVul\\\scriptsize test \(n=\num{428}\)}}                                                                                                                                                                                                                                                               \\
		\cmidrule(lr){3-5}\cmidrule(lr){6-8}\cmidrule(lr){9-11}
		                                                                                                                                                                                                                                                                                                                &                                       & \bfseries EM                                                                             & \bfseries \BLEUw                                                           & \bfseries \CodeBLEU                                                           & \bfseries EM                              & \bfseries \BLEUw                          & \bfseries \CodeBLEU                       & \bfseries EM                    & \bfseries \BLEUw                          & \bfseries \CodeBLEU                       \\
		\midrule
		\multirow{2}{*}{\makecell[c]{AVR                                                                                                                                                                                                                                                                     \\models}} & \cellcolor{lightyellow!50}\GraphTFive & \cellcolor{lightyellow!50}\bfseries 23.06                                                & \cellcolor{lightyellow!50}\bfseries 43.14                                  & \cellcolor{lightyellow!50}\bfseries 48.80                                     & \cellcolor{lightyellow!50}\bfseries 19.71 & \cellcolor{lightyellow!50}\bfseries 36.89 & \cellcolor{lightyellow!50}\bfseries 45.09 & \cellcolor{lightyellow!50} 5.37 & \cellcolor{lightyellow!50}\bfseries 32.83 & \cellcolor{lightyellow!50}\bfseries 39.83 \\
		\noalign{\vskip 1pt}
		\cdashline{2-11}[0.4pt/1.5pt]
		\noalign{\vskip 1pt}
		                                                                                                                                                                                                                                                                                                                & \VulMaster                            & 20.92                                                                                    & 35.73                                                                      & 45.40                                                                         & 16.43                                     & 32.59                                     & 42.41                                     & 4.44                            & 26.28                                     & 36.04                                     \\
		\midrule
		\multirow{2}{*}{\makecell[c]{LLMs}}                                                                                                                                                                                                                                                                             & \GPTFourO                             & 5.28                                                                                     & 18.41                                                                      & 25.92                                                                         & 4.93                                      & 16.31                                     & 24.67                                     & 7.94                            & 20.24                                     & 28.20                                     \\
		\noalign{\vskip 1pt}
		\cdashline{2-11}[0.4pt/1.5pt]
		\noalign{\vskip 1pt}
		                                                                                                                                                                                                                                                                                                                & \GeminiTwoFivePro                     & 7.95                                                                                     & 31.08                                                                      & 37.54                                                                         & 4.93                                      & 26.88                                     & 33.98                                     & \bfseries 9.11                  & 32.02                                     & 37.32                                     \\
		\bottomrule
	\end{tabular}
	\label{tab:eva:model-comparison}
\end{table}

\subsection{Static Comparison (\texorpdfstring{RQ$_1$}{RQ1})}\label{sec:eval:baseline-comparison}

Table~\ref{tab:eva:model-comparison} shows that \GraphTFive outperforms \VulMaster on all three datasets and metrics. The relative \EM improvement is \pct{10.2} on \BigVul + \CVEfixes (\pct{23.06} vs. \pct{20.92}), \pct{20.0} on \ICVul (\pct{19.71} vs. \pct{16.43}), and \pct{21.1} on \PrimeVul (\pct{5.37} vs. \pct{4.44}).

\BLEUw and \CodeBLEU follow the same pattern. In table order, \BLEUw increases by \pct{20.7}, \pct{13.2}, and \pct{24.9}, and \CodeBLEU by \pct{7.5}, \pct{6.3}, and \pct{10.5}. %
To test whether these improvements are statistically significant, we ran paired tests on the combined C/C++ test set. We compared \EM outcomes with McNemar's exact test~\cite{mcnemar:1947:note,liddell:1976:practical} and per-sample \BLEUw and \CodeBLEU scores with two-sided Wilcoxon signed-rank tests~\cite{wilcoxon:1945:individual}. Because the test set is large, the Wilcoxon \(p\)-values use the normal approximation of the test statistic, corrected for tied scores. We use a significance level of \(\alpha = 0.05\). All tests yielded \(p \le 1.11 \times 10^{-3}\), so \GraphTFive's improvements over all three baselines on the combined test set are statistically significant. The caption of Table~\ref{tab:eva:model-comparison} lists the values per baseline.

The comparison with the closed \acp{llm} depends on the dataset and metric. On \BigVul + \CVEfixes and \ICVul, \GraphTFive outperforms both closed models on all three metrics. On \PrimeVul, \GeminiTwoFivePro achieves the highest \EM (\pct{9.11}), while \GraphTFive achieves the highest \BLEUw (\pct{32.83}) and \CodeBLEU (\pct{39.83}).

\begin{rqbox}[]{Answer to \rqOne}
	On the full C/C++ test sets, \GraphTFive outperforms \VulMaster on all three datasets and metrics, with relative \EM improvements of \pct{10.2} to \pct{21.1}. It also outperforms \GPTFourO and \GeminiTwoFivePro on \BigVul + \CVEfixes and \ICVul. On \PrimeVul, \GeminiTwoFivePro achieves the highest \EM, while \GraphTFive achieves the highest \BLEUw and \CodeBLEU.
\end{rqbox}%

\input{sections/case_study.tex}

\subsection{Number of Repair Hunks (\texorpdfstring{RQ$_2$}{RQ2})}\label{sec:eval:multi-hunk-comparison}

Single-hunk repairs make up \pct{42.3} of the test set. On these repairs, \GraphTFive achieves \pct{25.98} \EM vs. \VulMaster's \pct{22.94}, a relative improvement of \pct{13.25}. This subset is dominated by \CWE{119} (buffer overflows), \CWE{125} (out-of-bounds reads), and \CWE{476} (null pointer dereferences).

\begin{figure}[t]
	\centering
	\includegraphics[width=\textwidth]{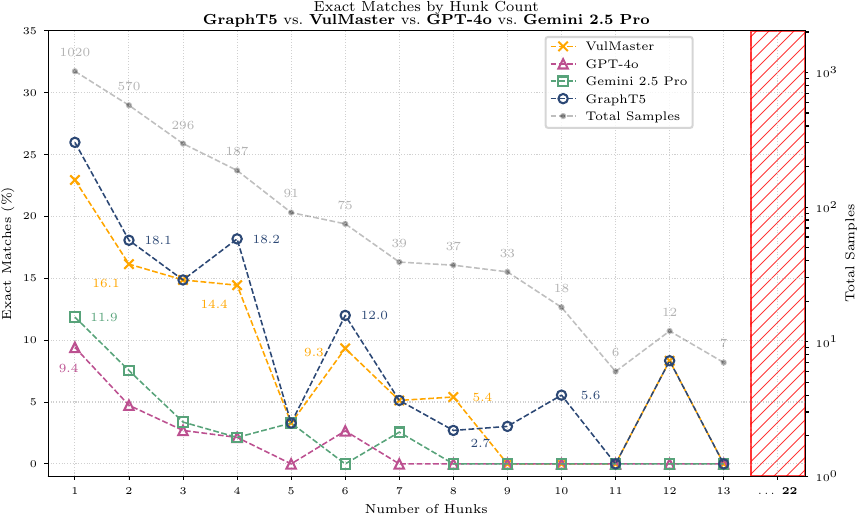}
	\caption{
		Exact match rates of \GraphTFive, \VulMaster, \GPTFourO, and \GeminiTwoFivePro by number of hunks in the target repair on the combined C/C++ test set. The gray curve shows the number of samples per count on the right logarithmic axis. The hatched bin covers counts from \num{14} to \num{22}, for which no model produces an exact match. High-count bins contain few samples and should be interpreted with caution.}
	\Description{Exact-match rates for GraphT5, VulMaster, GPT-4o, and Gemini 2.5 Pro are grouped by the number of hunks in the target repair. A gray curve shows the sample count on a logarithmic secondary axis.}
	\label{fig:eva:hunk-comparison-best-models}
\end{figure}

Figure~\ref{fig:eva:hunk-comparison-best-models} shows \EM by number of hunks in the target repair. \GraphTFive leads \VulMaster in the 2-, 4-, and 6-hunk bins and achieves a higher hunk-weighted \EM. \GPTFourO and \GeminiTwoFivePro do not exceed \GraphTFive or \VulMaster in any bin. Estimates for the sparse high-count bins are less stable.

The advantage is clearest in the well-populated bins. For 2-hunk repairs (\pct{23.6} of the test set), \GraphTFive achieves \pct{18.07} \EM vs. \VulMaster's \pct{16.14}, an \pct{11.96} relative improvement. The models tie at \num{3} hunks (\pct{14.86}), and at \num{4} hunks \GraphTFive reaches \pct{18.18} vs. \pct{14.44}, a \pct{25.93} relative improvement.

At \num{6} hunks, \GraphTFive achieves \pct{12.00} \EM vs. \VulMaster's \pct{9.33}, a \pct{28.57} relative improvement. \VulMaster leads in the sparse 8-hunk bin (\pct{5.41} vs. \pct{2.70}), while only \GraphTFive produces exact matches at \num{9} and \num{10} hunks. \GraphTFive and \VulMaster both produce an exact match at \num{12} hunks. \GPTFourO produces no exact match above \num{6} hunks, and \GeminiTwoFivePro none above \num{7} hunks.

\EMw summarizes performance across hunk counts by weighting each exact match by the number of hunks in the target repair (Section~\ref{sec:eval:research-questions}). On the three test sets combined, \GraphTFive achieves an \EMw of \pct{13.54} vs. \VulMaster's \pct{11.91}, a \pct{13.74} relative improvement. This advantage comes from gains in several common target sizes rather than from a uniform gain at every count. \GPTFourO and \GeminiTwoFivePro reach an \EMw of \pct{3.18} and \pct{4.33}. Their \EMw falls further below their \EM because a larger share of their exact matches are single-hunk repairs (\pct{70.1} and \pct{66.5} vs. \pct{57.1} for \GraphTFive).

\begin{rqbox}{Answer to \rqTwo}
	On single-hunk repairs, \GraphTFive improves \EM over \VulMaster from \pct{22.94} to \pct{25.98}, a \pct{13.25} relative gain. It also achieves a higher hunk-weighted \EM (\pct{13.54} vs. \pct{11.91}, a \pct{13.74} relative improvement), leads in the 2-, 4-, and 6-hunk bins, and is the only model with exact matches at \num{9} and \num{10} hunks. \GPTFourO and \GeminiTwoFivePro reach an \EMw of \pct{3.18} and \pct{4.33}.
\end{rqbox}

\subsection{Performance across CWE types (\texorpdfstring{RQ$_3$}{RQ3})}\label{sec:eval:cwe-wise-comparison}

\begin{figure}[H]
	\centering
	\includegraphics[width=\textwidth]{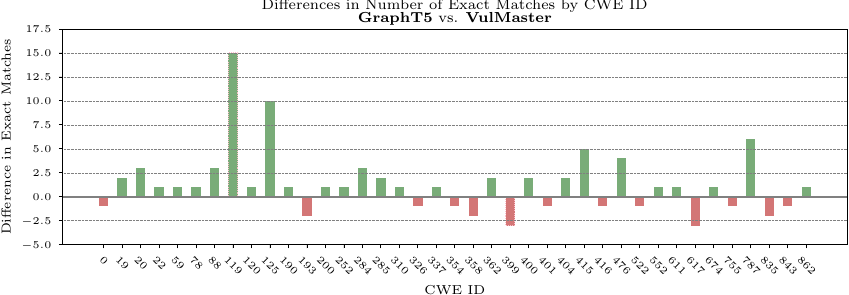}
	\caption{Difference in exact matches by \ac{cwe} category between \GraphTFive and \VulMaster on the combined C/C++ test set. Positive values indicate categories where \GraphTFive produces more exact matches; negative values indicate categories where \VulMaster produces more exact matches. Only categories with non-zero differences are shown.}
	\Description{Bars show GraphT5's exact-match count minus VulMaster's count in each weakness category with a nonzero difference. Positive bars favor GraphT5.}
	\label{fig:eva:comparison-best-models}
\end{figure}

Figure~\ref{fig:eva:comparison-best-models} shows the difference in exact matches per \ac{cwe}. \GraphTFive achieves its largest gains for buffer overflows (\CWE{119}, +15 exact matches), out-of-bounds reads (\CWE{125}, +10), out-of-bounds writes (\CWE{787}, +6), double frees (\CWE{415}, +5), and null pointer dereferences (\CWE{476}, +4). These five memory-safety categories account for \num{40} of the \num{72} additional exact matches in the categories where \GraphTFive leads. Across all \num{101} categories, \GraphTFive produces more exact matches in \num{26}, the same number in \num{62}, and fewer in \num{13}. Overall, it produces \num{52} more exact matches than \VulMaster (\num{464} vs. \num{412}).

Appendix Figure~\ref{app:fig:cwe-similarity-scores} reports the per-\ac{cwe} \CodeBLEU and \BLEUw scores. The categories with the largest exact-match gains generally also have higher \BLEUw and \CodeBLEU scores, particularly \CWE{119}, \CWE{125}, \CWE{787}, and \CWE{415}. Exact match may still undervalue functionally correct alternatives, because a vulnerability can often be fixed in several ways.

For \CWE{119}, the largest category with \num{405} samples, \GraphTFive produces \num{100} exact matches vs. \VulMaster's \num{85} (\pct{24.69} vs. \pct{20.99}). It also achieves higher \CodeBLEU (\pct{49.83} vs. \pct{44.09}) and \BLEUw (\pct{42.65} vs. \pct{33.54}). For \CWE{125}, \GraphTFive adds \num{10} exact matches (\pct{15.13} vs. \pct{10.92}) and improves \CodeBLEU from \pct{39.61} to \pct{45.41} and \BLEUw from \pct{28.61} to \pct{38.96}. These two categories account for \num{25} additional exact matches, and all three metrics agree.

The same pattern holds for other memory-safety categories. For \CWE{787}, \GraphTFive gains \num{6} exact matches (\pct{12.96} vs. \pct{9.26}), with higher \CodeBLEU (\pct{45.59} vs. \pct{42.50}) and \BLEUw (\pct{41.05} vs. \pct{33.22}). For \CWE{415}, it gains \num{5} exact matches (\pct{20.00} vs. \pct{5.71}), with \CodeBLEU rising from \pct{44.22} to \pct{52.00} and \BLEUw from \pct{36.71} to \pct{45.93}. \GraphTFive also gains \num{4} exact matches on \CWE{476} (\pct{15.20} vs. \pct{12.00}). These improvements are consistent with the value-flow and lifecycle relations that the \ac{vdg} represents, but per-category results alone do not establish causality.

Not every category follows the overall trend. \VulMaster leads by \num{3} exact matches for both \CWE{399} and \CWE{617}, and by \num{2} for \CWE{835}. In all three categories, however, \GraphTFive's \CodeBLEU is slightly higher (\pct{40.41} vs. \pct{39.80}, \pct{48.88} vs. \pct{47.04}, and \pct{46.52} vs. \pct{45.79}, respectively), so exact match and similarity can diverge. Both models achieve perfect \CodeBLEU and \BLEUw scores for \CWE{682} and \CWE{776}, but each of these categories contains only \num{2} test samples.

\begin{rqbox}{Answer to \rqThree}
	\GraphTFive's largest gains are in memory-safety categories: \CWE{119}, \CWE{125}, \CWE{787}, \CWE{415}, and \CWE{476} contribute \num{40} additional exact matches. Across all \num{101} categories, \GraphTFive leads in \num{26}, ties in \num{62}, and trails in \num{13}, with \num{52} more exact matches than \VulMaster overall. Results for small \ac{cwe} groups should be interpreted with caution.
\end{rqbox}

\begin{table}[H]
	\centering
	\caption{Compilation rate and pass@50 on the \num{81} single-file vulnerabilities of \datasetname{Vul4C} and on their single-function subset. Rates are percentages of the vulnerabilities in each set; counts are given in parentheses.}
	\small
	\setlength{\tabcolsep}{6pt}
	\sbox0{\begin{tabular}{@{}lcccc@{}}
			\toprule
			                                     & \multicolumn{2}{c}{Single-file (\num{81})} & \multicolumn{2}{c}{Single-function\textsuperscript{*} (\num{58})}                                                       \\
			\cmidrule(lr){2-3}\cmidrule(l){4-5}
			Model                                & Compilation rate                           & pass@50                                                           & Compilation rate         & pass@50                  \\
			\midrule
			\VRepair\textsuperscript{a}          & \pct{18.5} (15)                            & \pct{3.7} (3)                                                     & \pct{17.2} (10)          & \pct{5.2} (3)            \\
			\VulRepair\textsuperscript{a}        & \pct{34.6} (28)                            & \bfseries\pct{14.8} (12)                                          & \pct{37.9} (22)          & \bfseries\pct{19.0} (11) \\
			\VQM\textsuperscript{a}              & \pct{29.6} (24)                            & \pct{8.6} (7)                                                     & \pct{34.5} (20)          & \pct{12.1} (7)           \\
			\VulMaster\textsuperscript{a}        & \pct{49.4} (40)                            & \pct{11.1} (9)                                                    & \pct{51.7} (30)          & \pct{15.5} (9)           \\
			\rowcolor{lightyellow!50}\GraphTFive & \bfseries\pct{55.6} (45)                   & \bfseries\pct{14.8} (12)                                          & \bfseries\pct{67.2} (39) & \bfseries\pct{19.0} (11) \\
			\bottomrule
		\end{tabular}}
	\usebox0\par
	\begin{minipage}{\wd0}
		\scriptsize
		\textsuperscript{*}Developer fix changes exactly one function.\par
		\textsuperscript{a}Hu et al.~\cite{hu:2025:sok}: results derived from the detailed per-vulnerability outcomes.
	\end{minipage}
	\label{tab:eva:vul4c-comparison}
\end{table}

\subsection{Dynamic Validation on C/C++ (\texorpdfstring{RQ$_4$}{RQ4})}\label{subsec:eval:vul4c}

Table~\ref{tab:eva:vul4c-comparison} compares \GraphTFive with the four learning-based \ac{avr} models evaluated by Hu et al.~\cite{hu:2025:sok} on the \num{81} single-file vulnerabilities of \datasetname{Vul4C}.

\GraphTFive achieves the highest compilation rate (\num{45} vulnerabilities) and ties with \VulRepair for the highest pass@50 (\num{12} vulnerabilities).

Across the \num{81} single-file \datasetname{Vul4C} vulnerabilities, none of the five models produces a plausible patch for \num{56}. The failures concentrate in broader repairs: \num{34} of \num{42} multi-hunk cases have no plausible patch, compared with \num{6} of \num{16} single-hunk, single-line cases. Hu et al.~\cite{hu:2025:sok} restrict the learning-based models to single-file vulnerabilities, but these models repair one function per input. For a fix that changes two or more functions, each model in Table~\ref{tab:eva:vul4c-comparison} sees only one function at a time, without the other functions as context. In the evaluation of Hu et al., each baseline receives one of the changed functions. Single-file applicability therefore follows from how candidates are assembled, not from the models themselves. We therefore describe the scope of these models as single-function. The developer fixes of \num{23} of the \num{81} vulnerabilities (\pct{28.4}) change more than one function, so the baselines cannot reproduce these fixes. Only one of these \num{23} vulnerabilities receives a plausible patch from any model (\GraphTFive and \VulRepair); its developer fix changes one existing function and adds two helper functions.

Table~\ref{tab:eva:vul4c-comparison} therefore also reports results on the \num{58} vulnerabilities whose developer fix changes exactly one function. On this subset, pass@50 increases for all five models. \VulRepair and \GraphTFive both reach \pct{19.0} (\num{11} vulnerabilities each), \VulMaster \pct{15.5} (\num{9}), \VQM \pct{12.1} (\num{7}), and \VRepair \pct{5.2} (\num{3}). \GraphTFive also has the highest compilation rate on this subset (\pct{67.2}, \num{39} vulnerabilities).

\begin{rqbox}{Answer to \rqFour}
	On \datasetname{Vul4C}, a small benchmark of \num{81} single-file vulnerabilities, \GraphTFive produces compilable candidates for \num{45} and plausible patches for \num{12} (\pct{14.8} pass@50). It achieves the highest compilation rate and ties with \VulRepair for the highest pass@50.
\end{rqbox}

\begin{table}[H]
	\centering
	\caption{Test-passing patches on the \vulfourj-35 single-hunk benchmark. Rates are percentages of the \num{35} vulnerabilities; counts are given in parentheses. Results for external baselines are taken from Bui et al.~\cite{bui:saner2026:vulterminator} unless noted otherwise. Parameter counts are shown where public; n.p. marks non-public counts, and -- marks template-guided systems without a single comparable model size.}
	\scriptsize
	\setlength{\tabcolsep}{3pt}
	\begin{adjustbox}{max width=\textwidth}
		\begin{tabular}{@{}lcccccccccc@{}}
			\toprule
			                                   & \multicolumn{4}{c}{\makecell[c]{Frozen / Fine-Tuned\\LLMs}} & \multicolumn{2}{c}{LLMs}             & \multicolumn{2}{c}{\makecell[c]{Template-guided\\APR}} & \multicolumn{2}{c}{AVR-specific}                                                                                                                                                                                               \\
			\cmidrule(lr){2-5}\cmidrule(lr){6-7}\cmidrule(lr){8-9}\cmidrule(lr){10-11}
			                                   & \bfseries\Codex\textsuperscript{a}                          & \bfseries\Incoder\textsuperscript{a} & \bfseries\CodeGEN\textsuperscript{a}                   & \bfseries\CodeTFiveLarge\textsuperscript{a} & \bfseries\GPTFourO & \bfseries\GeminiTwoFivePro & \bfseries\NTR   & \bfseries\VulTerminator   & \bfseries\VulMasterLarge & \cellcolor{lightyellow!50}\bfseries\GraphTFive        \\
			\rowcolor{black!4}
			Parameters                         & \SI{12}{\billion}                                           & \SI{6}{\billion}                     & \SI{6}{\billion}                                       & \SI{770}{\million}                          & n.p.               & n.p.                       & --              & --                        & \SI{770}{\million}       & \cellcolor{lightyellow!50}\bfseries\SI{322}{\million} \\
			\midrule
			pass@10\textsuperscript{*} (total) & \pct{31.1} (10.9)                                           & \pct{25.7} (9)                       & \pct{22.9} (8)                                         & \pct{20.0} (7)                              & \pct{40.0} (14)    & \pct{48.6} (17)            & \pct{40.0} (14) & \bfseries \pct{54.3} (19) & \pct{25.7} (9)           & \cellcolor{lightyellow!50}\pct{42.9} (15)             \\
			\bottomrule
		\end{tabular}
	\end{adjustbox}
	\begin{minipage}{\textwidth}
		\scriptsize
		\textsuperscript{*}Ten candidates where applicable; \NTR uses up to \num{1000}. \VulTerminator validates candidates from all applicable templates until the first plausible patch; its beam size of \num{10} applies only to each masked-model prediction.\par
		\textsuperscript{a}Wu et al.~\cite{wu:issta2023:how}: fine-tuned variants except \Codex, whose count is averaged over \num{25} runs.\par
	\end{minipage}
	\label{tab:eva:vul4j-comparison}
\end{table}

\subsection{Java Adaptation (\texorpdfstring{RQ$_5$}{RQ5})}\label{subsec:eval:vul4j-benchmark-analysis}

This experiment tests whether the text-and-graph architecture transfers to Java under dynamic validation. Section~\ref{sec:eval:training-setup} describes the fine-tuning data. Table~\ref{tab:eva:vul4j-comparison} reports pass@10 on the \vulfourj-35 single-hunk benchmark, together with the baselines reported by Bui et al.~\cite{bui:saner2026:vulterminator}. \GraphTFive produces test-passing patches for \num{15} cases, one more than \GPTFourO and \NTR. Only \GeminiTwoFivePro (\num{17}) and \VulTerminator (\num{19}) report higher totals.

\GraphTFive has \SI{322}{\million} parameters, fewer than half of \VulMasterLarge's \SI{770}{\million} and far fewer than the multi-billion-parameter open code \acp{llm}. The parameter counts of \GPTFourO and \GeminiTwoFivePro are not public. \VulTerminator achieves the highest score, but its templates target Java repair patterns and would need language- and template-level adaptation for C/C++. Bui et al.~\cite{bui:saner2026:vulterminator} report that heuristic templates produce most of its successful repairs, so its result mainly reflects a Java-specific, template-guided pipeline. \GraphTFive, in contrast, uses the same learned architecture for C/C++ and Java, with language-specific fine-tuning.

\begin{rqbox}{Answer to \rqFive}
	After fine-tuning on Java vulnerability fixes, \GraphTFive produces test-passing patches for \num{15} of \num{35} \vulfourj vulnerabilities within ten candidates (\pct{42.9} pass@10). This is one more than the reported plausible totals of \GPTFourO and \NTR. Only \GeminiTwoFivePro (\num{17}) and \VulTerminator (\num{19}) report higher totals. \GeminiTwoFivePro is a general-purpose \ac{llm} of non-public size, and \VulTerminator relies on Java-specific repair templates.
\end{rqbox}

%% file: sections/case_study.tex
\subsubsection{Case Study}\label{sec:eval:case-study}

\begin{figure}[t]
	\centering
	\begin{tcolorbox}[colback=white!50!lightgray!30, arc=5pt,colframe=lightgray!20, boxrule=0pt, boxsep=0pt, left=12pt, right=0pt, top=0pt, bottom=0pt, width=0.96\textwidth]
\begin{minted}{cpp}
bool_t xdr_krb5_principal(XDR *xdrs, krb5_principal *objp) {
  int ret;
  char *p = NULL;
  krb5_principal pr = NULL;
  static krb5_context context = NULL;
  if (!context && kadm5_init_krb5_context(&context))
    return (FALSE);
  switch (xdrs->x_op) {
  case XDR_ENCODE:
    if (*objp) {
      if ((ret = krb5_unparse_name(context, *objp, &p)) != 0)
        return FALSE;
    }
    if (!xdr_nullstring(xdrs, &p))
      return FALSE;
    if (p)
      free(p);
    break;
  case XDR_DECODE:
    if (!xdr_nullstring(xdrs, &p))
      return FALSE;
    if (p) {
      ret = krb5_parse_name(context, p, &pr);
      if (ret != 0)
        return FALSE;
      *objp = pr;
      free(p);
    } else
      *objp = NULL;
    break;
  case XDR_FREE:
    if (*objp != NULL)
      krb5_free_principal(context, *objp); |\tikzmark{vultop}|
      |\textcolor{red}{<vul-start>}|break;|\textcolor{red}{<vul-end>}|
  }
  return TRUE;
}
\end{minted}
	\end{tcolorbox}
	\vspace*{1.85cm}%
	\caption{Case-study comparison of patches generated by \GraphTFive and \VulMaster for a double-free / use-after-free vulnerability (\CWE{415}/\CWE{416}) in the Kerberos stack.}
	\Description{A Kerberos function frees a principal without resetting its pointer. GraphT5 adds a null assignment after the free, whereas VulMaster inserts a null check before the free.}
	\label{fig:eva:case-study}

	\begin{tikzpicture}[overlay, remember picture]
		\tcbset{
			enhanced jigsaw,
			drop lifted shadow,
			colback=white,
			coltitle=black,
			colbacktitle=white!50!gray!50,
			opacitybacktitle=0.95,
			opacityback=1,
			colframe=white!75!black,
			boxrule=0.5pt,
			boxsep=0pt, left=0pt, right=0pt, top=0pt, bottom=0pt,
			toptitle=2pt,
			top=2pt,
			bottomtitle=2pt,
			halign title=center,
			fontupper=\scriptsize,
		}

		\coordinate (fix1) at ($(pic cs:minted-34-start) + (0,-\the\depth)$);
		\coordinate (fix2) at ($(pic cs:minted-33-end) + (0,-\depth+\height)$);

		\draw[path fading=north, fit fading=true] ($(fix1) + (0.125,0.125)$) -- (fix1);
		\draw[-Stealth] (fix1) -- ++(0,-1) |- ++(0.5,-1) node[right] {%
			\begin{minipage}{0.455\textwidth}
				\begin{tcolorbox}[title={\scriptsize{}Patch generated by \VulMaster.}]
\begin{minted}{cpp}
if (*objp != NULL)
    krb5_free_principal(context, *objp); 
else
    return FALSE;
break; 
\end{minted}
				\end{tcolorbox}
			\end{minipage}};
		\draw[path fading=south, fit fading=true] ($(fix2) + (-0.125,-0.125)$) -- (fix2);
		\draw[-Stealth] (fix2) -- ++(0,1) |- ++(-1,0.5) node[left] {%
			\begin{minipage}{0.455\textwidth}
				\begin{tcolorbox}[title={\scriptsize{}Patch generated by \GraphTFive.}]
\begin{minted}{cpp}
if (*objp != NULL)
    krb5_free_principal(context, *objp); 
*objp = NULL;
break; 
\end{minted}
				\end{tcolorbox}
			\end{minipage}};
	\end{tikzpicture}%
\end{figure}

To illustrate how \GraphTFive repairs vulnerabilities, we examine an example from the \BigVul + \CVEfixes test set. Figure~\ref{fig:eva:case-study} shows the sample and the patches generated by \GraphTFive and \VulMaster.

The vulnerable function belongs to MIT's Kerberos 5 authentication stack and runs in administrative daemons, which typically have elevated privileges.
The function operates in one of three modes (\cpp{XDR_ENCODE}, \cpp{XDR_DECODE}, \cpp{XDR_FREE}). In \cpp{XDR_FREE} mode, it releases the pointer \cpp{*objp} with \cpp{krb5_free_principal} but does not reset it.
A second call therefore finds a non-null stale pointer and frees it again, which corrupts the heap and terminates the process.
The stale pointer may also enable use-after-free behavior in privileged code, consistent with the high \ac{cvss} severity of this vulnerability~\cite{hudson:2015:github}.

\VulMaster wraps the free site in a conditional that returns \cpp{FALSE} if the pointer is already \cpp{NULL}. This resembles a common defensive pattern but has no effect here: on the failing execution path, the pointer is not null, so the new branch is never taken and the double free remains.

\GraphTFive instead sets the pointer to null after freeing the principal.
Any subsequent call then sees a null pointer and does not free the object again.
The \GraphTFive patch thus removes the stale-pointer state, whereas \VulMaster's patch changes the control flow without removing it.

The \ac{vdg} excerpt in Appendix Figure~\ref{app:fig:vdg-double-free} links the free site to source lines that use \texttt{*objp} and to their control-flow context. It supplies a structural view of the stale-pointer path alongside the sequential function input.

%% file: sections/discussion.tex
This section discusses four aspects of the results: how the decoder balances the two inputs, the constraints of function-level repair, the influence of dataset characteristics, and the limitations of the evaluation metrics.

\begin{figure}
	\centering
	\includegraphics[width=\textwidth]{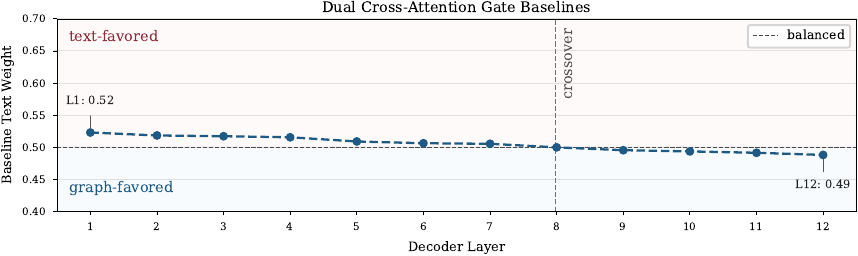}
	\caption{Learned gate baselines of the dual cross-attention after C/C++ pretraining. The plotted \(\sigma(\alpha^\ell)\) is the text-branch weight before the token-dependent adjustment; the graph branch receives its complement.}
	\Description{The learned baseline text weights across twelve decoder layers range from 0.488 to 0.523. The graph branch receives the complementary baseline weight.}
	\label{fig:discussion:cross-attention-gates}
\end{figure}

\subsection{Modality Balance in the Decoder}
\label{sec:discussion:gate-analysis}

A dual-input model could learn to ignore one of its inputs. We therefore analyze whether any decoder layer has a built-in preference for the text or the graph memory. Each decoder layer weights the two cross-attention outputs with a gate (Equation~\ref{seq:impl:eq:cross-attn}). The gate combines a learned per-layer baseline \(\sigma(\alpha^\ell)\), the text weight before any token-level adjustment, with a token-dependent term. The graph branch receives the complementary weight. The baselines thus show the global preference of each layer; the realized weights also depend on the token-level adjustment.

Figure~\ref{fig:discussion:cross-attention-gates} shows the learned baselines after C/C++ pretraining. The text baseline averages \num{0.519} in layers \num{1}--\num{4}, then declines steadily, crosses \num{0.5} between layers \num{7} and \num{8}, and reaches \num{0.488} at layer \num{12}. Across all layers, the text baseline stays within \num{0.488}--\num{0.523}, so the graph baseline is never below \num{0.477}. No decoder layer therefore has a strong global preference for either modality: each token-level gate starts from a nearly equal prior, with a slight shift from text in the lower layers toward the graph in the upper layers. The decoder thus does not discard either input: every layer starts from a balanced combination of the text and the graph memory.

\subsection{Function-Level Constraints and Inter-procedural Potential}
\label{sec:discussion-context}

We evaluate at the function level to allow a direct comparison with prior work~\cite{chen:tse2023:neural, fu:fse2022:vulrepair, zhou:icse2024:out}. This scope is a methodological choice, not an architectural limitation, but it constrains both the tasks the model can attempt and the supervision available during training. This has two consequences.

First, repair difficulty grows sharply with the number of hunks, even within a single function: \GraphTFive's exact-match accuracy falls from \pct{25.98} on single-hunk repairs to \pct{5.56} on 10-hunk repairs. Coordinating several edits within one function is therefore already difficult.

Second, current C/C++ datasets, including \BigVul + \CVEfixes, \ICVul, and \PrimeVul~\cite{zhou:icse2024:out, lu:2025:icvul, ding:icse2025:vulnerability}, treat each function changed in a vulnerability-fixing commit as an independent sample. This keeps the task tractable and comparable, but it discards which functions were changed together in the same fix. Repairs that span multiple functions are split across samples, which removes the caller-callee relations and the coordinated edits of the original commit.

The \ac{cpg} on which the \ac{vdg} is built also supports inter-procedural analysis. Connecting vulnerable sources to sinks across function boundaries would recover the context that function-level samples lose. Combined with long-context code models such as \modelname{Qwen3.8-27B}~\cite{qwenteam:2026:qwen38max}, inter-procedural \acp{vdg} could select cross-function dependencies while the text branch provides the surrounding source code. We plan to extend \GraphTFive to inter-procedural \acp{vdg} in future work.

\subsection{Dataset Characteristics Shape Performance}
\label{sec:discussion:evaluation-bias}

The \rqOne results suggest that the splitting method influences performance. \PrimeVul's chronological split yields the lowest scores (\EM of \pct{5.37} for \GraphTFive and \pct{4.44} for \VulMaster), which suggests temporal drift in vulnerability patterns. \PrimeVul is also the only dataset on which the closed \acp{llm} are competitive: \GeminiTwoFivePro achieves the highest \EM, while \GraphTFive leads on \BLEUw and \CodeBLEU. On \BigVul + \CVEfixes and \ICVul, both task-specific models clearly outperform the closed \acp{llm}.

The chronological split also changes the \ac{cwe} distribution between training and test data. \PrimeVul splits samples by commit date to prevent temporal leakage, for example semantically equivalent but syntactically different functions with the same vulnerability that appear in both the training and the test set. This is a reasonable mitigation, but we found that it strongly skews the \ac{cwe} distribution. For example, \CWE{787} (out-of-bounds write) makes up less than \pct{7} of the training split but more than \pct{16} of the test split. Similarly, \CWE{703} (improper handling of exceptional conditions) makes up more than \pct{10} of the test split but less than \pct{4} of the training split. Most \acp{cwe} in \PrimeVul show such a shift.

The relative \EM advantage over \VulMaster is \pct{10.2} on \BigVul + \CVEfixes, \pct{20.0} on \ICVul, and \pct{21.1} on \PrimeVul. The larger gains on \ICVul and \PrimeVul are consistent with a benefit from stronger dataset curation. One possible explanation is imprecise localization in tangled commits. A tangled commit mixes the fix with unrelated changes, so the marked vulnerable lines include lines unrelated to the vulnerability. Slicing from these lines adds unrelated dependencies to the \ac{vdg}. Cleaner samples mark fewer unrelated lines and therefore align the graph better with the vulnerability. A controlled study is needed to test this explanation.

\subsection{Metrics Reveal Evaluation Limitations}
\label{sec:discussion_evaluation}

\EM measures a different property than \BLEUw and \CodeBLEU. For \CWE{617}, \GraphTFive produces \num{3} fewer exact matches than \VulMaster but achieves higher \CodeBLEU (\pct{48.88} vs. \pct{47.04}) and \BLEUw (\pct{43.06} vs. \pct{40.47}). On \PrimeVul, the closed \acp{llm} achieve higher \EM, while \GraphTFive leads on \BLEUw and \CodeBLEU. A prediction can share many tokens with the reference but miss the decisive security check; conversely, a correct alternative fix may differ substantially from the developer patch. Similarity alone cannot distinguish these cases.

Dynamic validation gives a different picture. On \datasetname{Vul4C}, \GraphTFive produces plausible patches for \num{12} of \num{81} vulnerabilities, three more than \VulMaster but the same number as \VulRepair. On \vulfourj, it produces test-passing patches for \num{15} of \num{35} cases within ten candidates, behind \GeminiTwoFivePro and \VulTerminator. Plausible patches pass the available exploits and tests but can still overfit them~\cite{wu:issta2023:how}, so test plausibility is an upper bound on correct repair. These benchmarks show that the architecture can be adapted to another language; they do not show a general advantage in functional correctness.

Static similarity metrics are nevertheless still needed in the current \ac{avr} setting. To the best of our knowledge, no published C/C++ vulnerability-repair benchmark provides executable \ac{pov} tests at the scale of \BigVul, \CVEfixes, \ICVul, or \PrimeVul. As in prior learning-based \ac{avr} work~\cite{harer:neurips2018:learning,zhou:emse2022:spvf,chi:tse2022:seqtrans,fu:fse2022:vulrepair,chen:tse2023:neural,fu:tosem2024:vision,zhou:icse2024:out,de-fitero-dominguez:eaai2024:enhanced,wen:2025:vulr2,wang:ase2026:spvr}, \EM, \BLEUw, and \CodeBLEU therefore remain the standard large-scale substitute. They should be read as approximations, not as direct evidence of vulnerability removal. The syntax and data-flow components of \CodeBLEU assume parseable programs. On instruction sequences, they partly reduce to weighted n-gram matching, as in prior \ac{avr} evaluations that use the same representation~\cite{chen:tse2023:neural,zhou:icse2024:out}. Recent metrics such as \CodeBERTScore~\cite{zhou:emnlp2023:codebertscore} offer richer similarity measures within the same paradigm, but large-scale dynamic frameworks that directly measure vulnerability removal are still missing.

%% file: sections/validity.tex
\paragraph{Evaluation Metrics.} \EM can penalize correct patches that differ from the developer patch, while \BLEUw and \CodeBLEU can reward similar patches that do not remove the vulnerability. These metrics therefore indicate vulnerability removal only indirectly. To mitigate this threat, we also evaluate \GraphTFive on \datasetname{Vul4C} and \vulfourj, whose exploits, regression tests, and \ac{pov} tests check whether a patch removes the vulnerability. These benchmarks are small (\num{81} and \num{35} vulnerabilities), but they provide dynamic evidence of vulnerability removal.

\paragraph{Dataset Consistency.} For C/C++, \GraphTFive and \VulMaster use the same splits of \BigVul + \CVEfixes, \ICVul, and \PrimeVul. This allows a direct comparison but does not guarantee that training and test records are independent. \PrimeVul reduces this risk with a chronological split, which places older fixes in the training data and newer fixes in the test data. This split, however, shifts the \ac{cwe} distribution between training and test data (Section~\ref{sec:discussion:evaluation-bias}). For Java, we fine-tune on vulnerability fixes from the corpus of Bui et al.~\cite{bui:saner2026:vulterminator}, which excludes the commits of their test datasets, including \vulfourj (Section~\ref{sec:eval:training-setup}). No fine-tuning example therefore comes from a fix commit of a \vulfourj test vulnerability.

\paragraph{C/C++ Preprocessor Handling.} C/C++ preprocessor directives include or exclude code depending on compiler flags. %
Because functions are processed in isolation, without project-specific build configurations, code in inactive preprocessor blocks could be missing from the \ac{cpg} and thus from the \ac{vdg}. %
We therefore set all preprocessor conditions to true so that most code paths are included. This heuristic may not match the actual build configuration, especially in complex projects such as the Linux kernel, so some vulnerable code inside conditional blocks could still be missed.
Because line breaks are syntactically significant in preprocessor directives, we insert special line-break tokens during \ac{vdg} construction only; the text input of the model is unchanged.

\paragraph{Data Quality and Leakage.} Commit-based datasets inherently contain noise~\cite{yadav:acsac2024:rr}, such as tangled commits that mix unrelated changes with the vulnerability fix~\cite{herzig:msr2013:impact}.
Such noise can mislabel fix-relevant lines and affects the two representations differently: graph slicing depends on the marked regions, whereas the text encoder sees the whole function.
A separate concern is leakage between the pretraining corpus of the base model and the fine-tuning and evaluation datasets.
\GraphTFive and \VulMaster both build on \CodeTFive, whose exact training set is not public.
We therefore cross-referenced the commits of the three C/C++ datasets (\BigVul + \CVEfixes: \num{976} commits, \ICVul: \num{4372} commits, \PrimeVul: \num{6872} commits) with the documented source of \CodeTFive's training data, a subset of Google's BigQuery GitHub dataset~\cite{google_bigquery_github_dataset}.
Some of these commits occur in the BigQuery dataset (\SI{265}{\million} commits in total), but the subset used to train \CodeTFive (\SI{1}{\million} C samples) is not available, so we cannot quantify the overlap with the \CodeTFive pretraining data.
Any such leakage would mainly affect the \CodeTFive-based models, \GraphTFive and \VulMaster.
For the closed \acp{llm}, the training data are not disclosed, so we cannot audit benchmark exposure for \GPTFourO and \GeminiTwoFivePro. Their knowledge cutoffs are October 2023~\cite{openai:2024:gpt4o} and January 2025~\cite{googlecloud:2025:gemini}, and many fixing commits in \BigVul, \CVEfixes, \ICVul, \PrimeVul, and \vulfourj are older. Code \acp{llm} are commonly trained on large crawls of public GitHub repositories~\cite{chen:2021:evaluating}. Both models may therefore have seen the vulnerable and the fixed code of our test samples, which would inflate their results. Such exposure favors the closed \acp{llm} and therefore does not weaken \GraphTFive's advantage over them.

\paragraph{Language Generalizability.} The \ac{vdg} relies on \Joern and its language-specific frontends. The results on C/C++ and Java suggest that the approach is not limited to one language family. The Java result, however, reflects cross-language adaptation rather than zero-shot transfer, because it includes fine-tuning on Java vulnerability fixes. In principle, \GraphTFive applies to any language with a suitable \ac{cpg} frontend, but we have validated it only on these two languages. Its performance on languages with substantially different paradigms, such as dynamic or functional languages, remains an open question.

\paragraph{Fault Localization.} Following the standard setting of learning-based \ac{avr}~\cite{zhou:emse2022:spvf, chen:tse2023:neural, fu:fse2022:vulrepair, zhou:icse2024:out}, our evaluation assumes perfect line-level fault localization (Section~\ref{sec:grapht5-overview}). This assumption reflects the scope of the repair task, not a simplification of our setup. In \GraphTFive, these annotations set the text markers, seed the \ac{vdg} slice, and assign the vulnerability roles. Localization errors can therefore change the graph content and its encoding, not only the marked positions in the text input. We have not measured this sensitivity. Deployment with predicted locations requires evaluating the complete detection-and-repair pipeline, which is outside the scope of this work.

\paragraph{Dataset Representativeness.}
Most datasets in this study fragment multi-function fixes, and some test sets, such as the dynamic validation benchmarks and many \ac{cwe} subsets, are small.
Generalization to vulnerabilities that differ substantially from those in the training data is not guaranteed.

\paragraph{Experimental Environment and Reproducibility.} The models we trained were evaluated under controlled hardware and software conditions, with fixed random seeds, deterministic PyTorch~\cite{pytorchteam:2025:pytorch} settings where supported, and the cuBLAS reproducibility guidelines~\cite{nvidiacorporation:2025:cublas}. We matched hyperparameters where this was meaningful for direct comparison and kept architecture-specific parameters as each model requires. These steps reduce variance but cannot eliminate all nondeterminism.

%% file: sections/related_work.tex
This section relates \GraphTFive to prior work on neural vulnerability repair and structure-aware code models.

\subsection{Sequence-to-Sequence Models for Vulnerability Repair}
\label{sec:related-seq2seq}

Most neural \ac{avr} models treat repair as \ac{s2s} translation from a vulnerable function to its repaired version~\cite{chen:tse2023:neural,fu:fse2022:vulrepair,zhou:icse2024:out}. Prior work has extended this text-only formulation with pretraining, structural information, and attention guidance.

\VRepair~\cite{chen:tse2023:neural} was one of the first Transformer-based \ac{avr} models. It translates vulnerable code into fixed code, pretrains on a large corpus of general bug-fix commits, and fine-tunes on a smaller vulnerability-fix dataset to compensate for the scarcity of labeled vulnerability data. Its word-level tokenization with a copy mechanism for rare tokens, however, limits its ability to introduce new tokens, and it represents code only as a linear token sequence. \SeqTrans~\cite{chi:tse2022:seqtrans} adds data-flow heuristics that move the tokens of vulnerable variable definitions and uses closer together in the input, and it normalizes variable names to reduce out-of-vocabulary tokens. \VulRepair~\cite{fu:fse2022:vulrepair} replaces training from scratch with \TFive-style pretraining on code and uses \ac{bpe} and relative position embeddings to better handle identifiers and token order.

A second line of work linearizes structural information into the input token stream. \SPVF~\cite{zhou:emse2022:spvf} represents each input by its serialized \ac{ast} together with a description of the vulnerability type or condition, and combines both through a pointer-generator network that can copy from the \ac{ast} and the code. Because the \ac{ast} is linearized, the model cannot pass messages along graph edges and must recover the tree structure implicitly through self-attention. \VulMaster~\cite{zhou:icse2024:out} combines this idea with a \CodeTFive model that it further pretrains on \VRepair's bug-fix corpus, uses the \ac{fid} architecture~\cite{izacard:eacl2021:leveraging} to process complete functions, and adds \ac{cwe} information as domain knowledge. Like \VRepair and \GraphTFive, it generates repairs in the token-level modification grammar. For tree serialization, it builds on the structure-preserving traversal of Hu et al.~\cite{hu:icpc2018:deep}. Like prior \ac{avr} models, \SPVF and \VulMaster assume perfect fault localization through marked vulnerable regions.

A third line of work keeps the input textual but guides attention toward vulnerable regions. \VQM~\cite{fu:tosem2024:vision}, inspired by Vision Transformers, introduces vulnerability queries that the decoder matches against the encoded code tokens, together with a learned vulnerability mask that amplifies attention on vulnerable locations in both encoder self-attention and decoder cross-attention. \VulAdvisor~\cite{zhang:ase2024:vuladvisor} instead generates natural-language repair suggestions for developers or downstream \ac{avr} tools. It fine-tunes \CodeTFive on pairs of vulnerable functions and suggestions derived from historical fixes and adds a repair-action loss that focuses the model on relevant constructs.

Outside the \ac{s2s} framing, \VulTerminator~\cite{bui:saner2026:vulterminator} returns to template-based \ac{apr} for Java vulnerability repair. It combines six heuristic templates for recurring vulnerability classes, such as XML External Entity injection and path traversal, with five data-driven templates that generate masked patch skeletons, whose missing parts a fine-tuned masked language model fills in. Bui et al. report that the heuristic templates produce most of the successful repairs on \vulfourjplus and \vulfourjl.

Inconsistent evaluation setups complicate comparisons between these models. The original \VQM study did not compare against \VulMaster; a later evaluation by Wang et al.~\cite{wang:tosem2026:evaluation} on the \REEF benchmark~\cite{wang:2023:reef} found that \VulMaster outperforms \VQM and the other tested \ac{avr} models, with \pct{28.94} \EM. Yadav et al.~\cite{yadav:acsac2024:rr} also found train-test contamination in the splits of \VQM's original evaluation. Han et al.~\cite{han:icse2026:rethinking} extend this concern to fine-tuned \ac{avr} in general: random splits can overstate repair performance through train-test overlap, and semantics-preserving transformations show that models rely on spurious surface features.

In summary, prior \ac{avr} models keep code as plain text (\VRepair, \SeqTrans, \VulRepair, \VQM, \VulAdvisor), linearize structure into a token sequence (\SPVF, \VulMaster), or rely on fixed templates (\VulTerminator). \GraphTFive instead consumes a vulnerability-focused graph natively, alongside the source code, with two encoders fused through dual cross-attention.

\subsection{Graph Representations for Vulnerability Detection and Repair}

Graph representations of code are widely used for vulnerability detection but rarely for repair.

For detection, \Devign~\cite{zhou:neurips2019:devign} applies \acp{ggnn} to \acp{cpg}, the unified graph representation introduced by Yamaguchi et al.~\cite{yamaguchi:sp2014:modeling}. \FUNDED~\cite{wang:tifs2020:combining} applies multi-relational \acp{gnn} to flow-sensitive graphs. \SySeVR~\cite{li:tdsc2022:sysevr} performs interprocedural forward and backward slicing from syntax-based vulnerability candidates and then linearizes the slices into token sequences. \SlicedLocator~\cite{wu:cose2023:slicedlocator} keeps a sliced graph, the \ac{sdg}, which it builds from \ac{coi} nodes with control-, data-, and call-dependence edges for statement-level localization. \VulLMGNN~\cite{liu:arxiv2024:vullmgnns} instead keeps full \acp{cpg} and fuses them with features from a pretrained code language model. These approaches show that graph representations are effective for classification, but they output a label or a score, not a repair.

For repair, graphs have been used in three ways. \Hoppity~\cite{dinella:iclr2020:hoppity} addresses general program repair with a message-passing network that edits \acp{ast} directly, for example by inserting or deleting nodes, and therefore needs no alignment between modalities. \APPATCH~\cite{nong:uss2025:appatch} addresses vulnerability repair by extracting a code slice from an \ac{sdg} through union slicing, based on \SySeVR, but passes the slice to an \ac{llm} as code rather than as a graph. Within \ac{avr}, \SPVF and \VulMaster include linearized syntax trees in their input, but their models operate on token sequences, not on the graph.

This body of work has two limitations for repair. First, when a graph is reduced to a token stream before it reaches the model, as in \SPVF, \VulMaster, and \APPATCH, the model can no longer reason along graph edges. Second, syntax-only representations may miss relations between distant but data-dependent code elements. \GraphTFive addresses both: a dedicated encoder consumes the \ac{vdg} as a graph, and its edges encode control-flow, control-dependence, data-dependence, and declaration-use relations that sequential representations do not preserve directly. To the best of our knowledge, \GraphTFive is the first \ac{avr} model to use a vulnerability-focused graph natively for patch generation.

\subsection{Structure-Aware Neural Code Models}

Beyond \ac{avr}, a broad body of work studies how to preserve code structure in neural models. Standard sequence models flatten code into tokens and lose its hierarchy and long-distance dependencies. Three architectural strategies address this problem.

The first strategy transforms the input. Hu et al.~\cite{hu:icpc2018:deep} feed \acp{ast} serialized with \ac{sbt} into an \ac{rnn} for code summarization and improve comment generation over plain token input. Sun et al.~\cite{sun:aaai2020:treegen} propose \TreeGen, a code-generation model with a dedicated \ac{ast} encoder and a decoder that generates \ac{ast} grammar rules, which yields syntactically valid outputs and mitigates long-range dependency problems.

The second strategy injects structure into the attention computation of a Transformer. Hellendoorn et al.~\cite{hellendoorn:iclr2019:global} add relative positional encodings derived from \acp{pdg}, which brings data flow into self-attention. \Graphormer~\cite{ying:neurips2021:transformers} biases attention scores by node adjacency, distance, and centrality and achieves state-of-the-art results on molecular graph benchmarks. \StructCoder~\cite{tipirneni:tkdd2024:structcoder} and \CSATrans~\cite{oh:arxiv2024:csatrans} apply similar ideas to code generation and add \ac{ast} and \ac{dfg} links as positional encodings and attention heads. These models show that a single Transformer can integrate structural relations in its attention or embedding layers without losing the global context of sequence models.

The third strategy uses separate encoders for separate modalities and fuses their outputs during decoding. LeClair et al.~\cite{leclair:icpc2020:improved} apply this strategy to code summarization: one encoder processes the source code as text, a graph neural network encodes the \ac{ast}, and the two outputs are fused before the summary is generated. To the best of our knowledge, dual-encoder models have not been applied to \ac{avr}, nor combined with code representations richer than \acp{ast}. \GraphTFive applies this strategy to \ac{avr} with a \CodeTFive-initialized text encoder and a dedicated graph encoder over the \ac{vdg}, fused in the decoder through gated dual cross-attention.

%% file: sections/conclusion.tex
We presented \GraphTFive, a dual-encoder repair model that processes source code and vulnerability-focused graphs as separate inputs. On three C/C++ datasets, it achieves higher scores than \VulMaster on all three static similarity metrics. Its hunk-weighted \EM reaches \pct{13.54} compared with \pct{11.91}, a \pct{13.74} relative improvement.

On \datasetname{Vul4C}, \GraphTFive produces plausible patches for \num{12} of \num{81} vulnerabilities, tied with \VulRepair for the best result. After fine-tuning on Java, it produces test-passing patches for \num{15} of \num{35} \vulfourj cases within ten candidates. These dynamic validation results show that the architecture can be adapted across languages. They also show that gains on static similarity metrics translate only partly into test-passing patches.

The results motivate repair models that use structural context alongside source text. Larger benchmarks for dynamic validation and semantic assessment are needed to measure such gains beyond static similarity metrics.

%% file: sections/data_availability.tex
We provide a replication package on \href{https://github.com/tuhh-softsec/GraphT5}{GitHub}.
It contains all datasets with the train/validation/test splits used in our experiments, the preprocessing scripts, and the model, training, and inference code with its configuration files.

%% file: sections/appendix_prestudy_vocabulary.tex
\begin{table}[ht]
	\centering
	\scriptsize
	\caption{Controlled vocabulary for the pre-study comparison in Table~\ref{tab:prestudy-avr}.}
	\label{tab:prestudy-vocab}
	\noindent\begin{minipage}[t]{0.485\linewidth}
		\vspace{0pt}
		\begin{tblr}{
				width = \linewidth,
				colspec = {Q[l,t,wd=1.7cm] X[l,t]},
				row{1} = {font=\bfseries, bg=prestudyHeader},
				row{2,21,26} = {font=\bfseries, bg=prestudyGroup},
				hlines = {white},
				hline{1,2,29} = {0.8pt, black},
				hline{21,26} = {0.4pt, gray!40},
				colsep = 4pt,
				rowsep = 0.5pt,
			}
			Acronym  & Description                                                   \\
			\SetCell[c=2]{l} Architecture                                            \\
			TR-E/D   & Transformer encoder--decoder                                  \\
			TR-DO    & Decoder-only Transformer                                      \\
			PG/Attn. & Pointer-generator network with attention                      \\
			GNN      & Graph neural network                                          \\
			TPL      & Template-based                                                \\
			PAT      & Pattern/example-based                                         \\
			MLM      & Masked language model                                         \\
			HYB      & Hybrid pipeline                                               \\
			GAN      & Generative adversarial network (NMT generator)                \\
			SYMB     & Symbolic or concolic execution                                \\
			SYNTH    & Property- or constraint-guided patch synthesis                \\
			INVAR    & Inductive invariant inference                                 \\
			CoT      & Chain-of-thought prompting                                    \\
			ToT      & Tree-of-thought prompting                                     \\
			AGENT    & Autonomous agent loop with tool use                           \\
			FiD      & Fusion-in-Decoder                                             \\
			(FT)     & Fine-tuned decoder-only model                                 \\
			RL       & Reinforcement-learning post-training                          \\
			\SetCell[c=2]{l} Localization                                            \\
			PFL-L    & Perfect line-level fault localization                         \\
			PFL-Func & Vulnerable function is given                                  \\
			E2E      & Part of the approach; no vulnerability witness available      \\
			PoC-L    & Derived from a required \ac{poc} exploit or sanitizer trace \\
			\SetCell[c=2]{l} Symbols                                                 \\
			\avrNR   & Not reported                                                  \\
			\avrNA   & Not applicable                                                \\
		\end{tblr}
	\end{minipage}\hfill
	\begin{minipage}[t]{0.485\linewidth}
		\vspace{0pt}
		\begin{tblr}{
				width = \linewidth,
				colspec = {Q[l,t,wd=1.7cm] X[l,t]},
				row{1} = {font=\bfseries, bg=prestudyHeader},
				row{2,20,28} = {font=\bfseries, bg=prestudyGroup},
				hlines = {white},
				hline{1,2,36} = {0.8pt, black},
				hline{20,28} = {0.4pt, gray!40},
				colsep = 4pt,
				rowsep = 0.5pt,
			}
			Acronym           & Description                                                          \\
			\SetCell[c=2]{l} Input representation                                                    \\
			TXT               & Source-token text                                                    \\
			LOC               & Marked vulnerable location                                           \\
			CWE               & Common Weakness Enumeration                                          \\
			AST-SEQ           & Serialized abstract-syntax-tree sequence                             \\
			DF                & Data-flow heuristics                                                 \\
			PROP              & Security property                                                    \\
			VQ                & Vulnerability query or query mask                                    \\
			NL                & Natural-language suggestion or description                           \\
			PROMPT            & Prompt-only input                                                    \\
			ICL               & In-context examples                                                  \\
			SLICE             & Static-analysis-derived slice                                        \\
			CPG               & Code property graph or CPG-derived graph input                       \\
			FUSE              & Learned multimodal fusion                                            \\
			POC               & Witnessing exploit input                                             \\
			SAN               & Sanitizer report or crash log                                        \\
			FDBK              & Validation feedback fed back into the prompt                         \\
			VIC               & Vulnerability-inducing-commit context                                \\
			\SetCell[c=2]{l} Repair scope                                                            \\
			Statement         & Generated edit is confined to one localized statement                \\
			Function          & Generated edits are confined to one function or method               \\
			File              & Generated edits are confined to one source file                      \\
			Cross-function    & Generated edits may span more than one function                      \\
			Multi-location    & Fixes edit several locations; the source does not report whether they form one or several hunks \\
			Single-hunk       & Every evaluated fix is one contiguous block of changed lines (hunk)  \\
			Single/multi-hunk & Fixes with one and with several hunks are explicitly evaluated       \\
			\SetCell[c=2]{l} Evaluation oracle                                                       \\
			SIM               & Static similarity metrics                                            \\
			SAST              & Static-analysis query or warning                                     \\
			COMP              & Compilation checks                                                   \\
			PoV               & Benchmark \ac{pov} (vulnerability-witnessing) tests                             \\
			EXP               & Replay of a concrete vulnerability-triggering input                  \\
			TEST              & Regression, unit, or functional tests                                \\
			MAN               & Manual validation                                                    \\
		\end{tblr}
	\end{minipage}
\end{table}

%% file: sections/appendix_llm_prompt.tex
\begin{center}
	\centering
	\begin{tcblisting}{
			enhanced,
			width=\linewidth,
			colback=white!98!black,
			colframe=lightgray!55!black,
			colbacktitle=white!94!black,
			coltitle=black,
			fonttitle=\bfseries\small,
			title={Prompt Template Used for Closed LLM Evaluation},
			boxrule=0.35pt,
			arc=3pt,
			left=4pt,
			right=5pt,
			top=4pt,
			bottom=5pt,
			listing only,
			listing options={
					basicstyle=\ttfamily\scriptsize,
					breaklines=true,
					columns=fullflexible,
					keepspaces=true,
					showstringspaces=false,
					xleftmargin=-8em
				},
			overlay unbroken={
					\node[
						fill=lightyellow!55,
						draw=black!20,
						rounded corners=2pt,
						font=\scriptsize\bfseries,
						anchor=north east,
						xshift=-6pt,
						yshift=-6pt
					] at (frame.north east) {zero-shot, deterministic};
				}
		}
		You are an expert in vulnerability repair. Your job is to fix the following function, subject to CWE-<ID>.
		Vulnerable lines have been marked with /*<vul-start>*/ ... /*<vul-end>*/.

		# Output Guidelines
		- Return exactly one fenced code block labeled repair.
		- Put the full repaired function inside the repair block.
		- Do not include explanations or any text before or after the repair block.

		## Required Output Format
		```repair
		<full repaired function>
		```

		## Vulnerable Function
		<annotated vulnerable function>
	\end{tcblisting}
	\captionsetup{hypcap=false}
	\captionof{figure}{Prompt template used for the closed-\ac{llm} comparison on the C/C++ datasets. The placeholder \texttt{CWE-<ID>} was replaced with the sample's vulnerability type, and the full vulnerable function was provided with inline vulnerability markers.}
	\label{fig:eva:llm-prompt}
\end{center}

%% file: sections/appendix_per_cwe_similarity.tex
Figure~\ref{app:fig:cwe-similarity-scores} reports the per-\ac{cwe} \BLEUw and \CodeBLEU scores for the categories with non-zero exact-match differences in Figure~\ref{fig:eva:comparison-best-models}.

\begin{center}
	\centering
	\includegraphics[width=\textwidth,height=0.68\textheight,keepaspectratio]{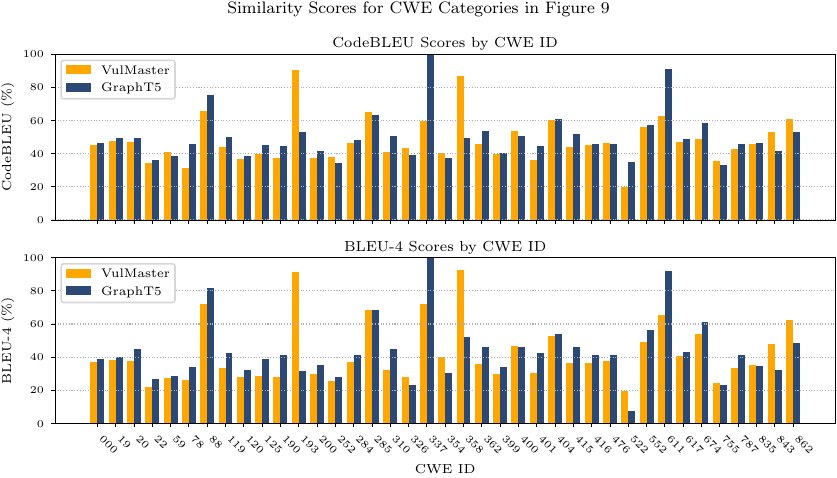}
	\captionsetup{hypcap=false}
	\captionof{figure}{\CodeBLEU and \BLEUw scores for the \ac{cwe} categories shown in Figure~\ref{fig:eva:comparison-best-models}. Scores are computed within each category rather than over the full test set.}
	\label{app:fig:cwe-similarity-scores}
\end{center}

%% file: sections/appendix_case_study_vdg.tex
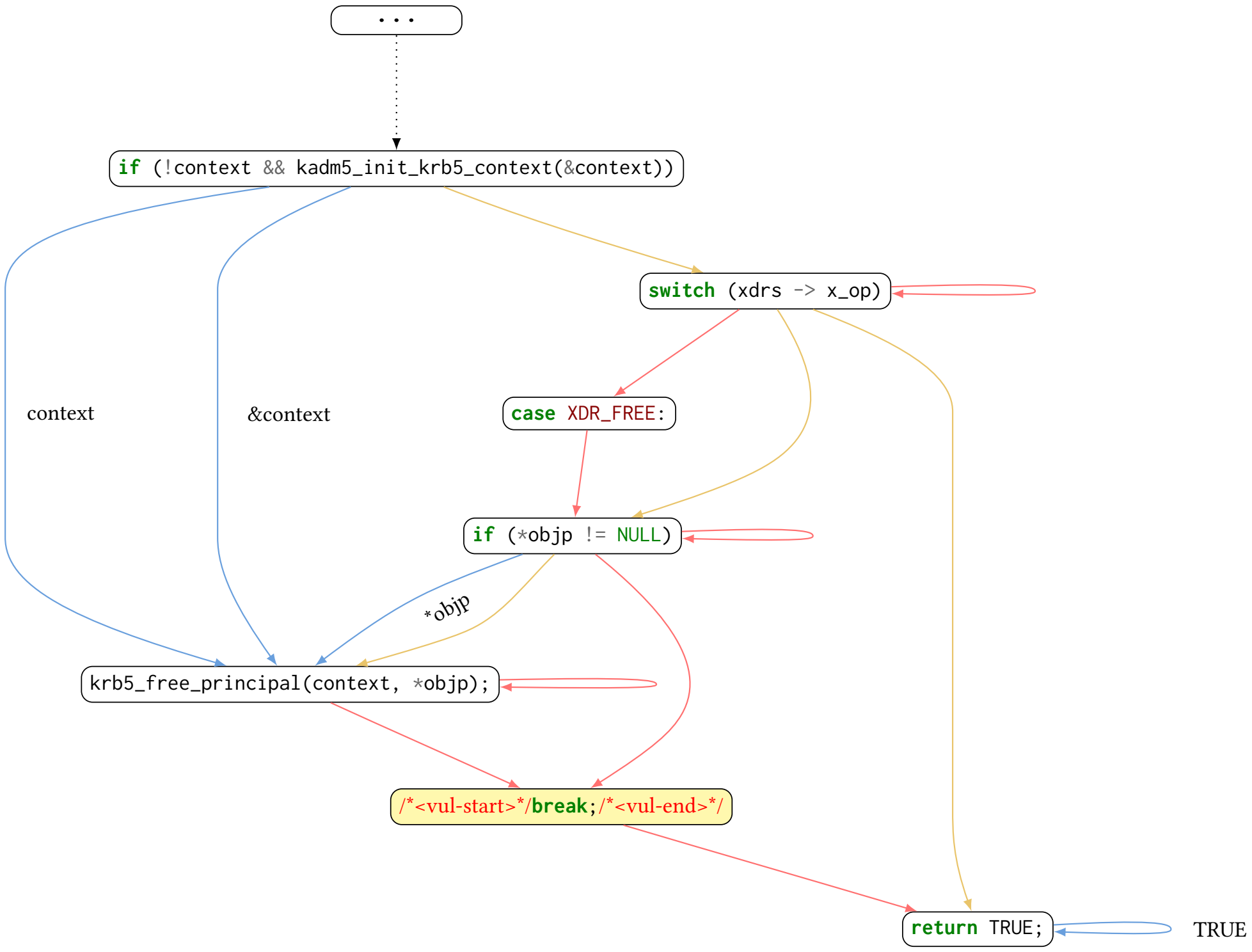
\begin{figure}[h]
	\centering
	\adjustbox{max width=0.84\textwidth, max height=0.52\textheight, keepaspectratio, trim=0 0 15pt 0}{
		\input{figures/evaluation/case_study/cwe_415_vdg_part.tex}
	}
	\caption{Excerpt from the \ac{vdg} of the C function containing a double-free (\CWE{415}) vulnerability.}
	\Description{A graph excerpt links the principal pointer and the statements relevant to freeing it, illustrating the context of the double-free vulnerability.}
	\label{app:fig:vdg-double-free}
\end{figure}

Figure~\ref{app:fig:vdg-double-free} shows the \ac{vdg} excerpt of the double-free sample discussed in Section~\ref{sec:eval:case-study}.

%% file: figures/evaluation/case_study/cwe_415_vdg_part.tex
\begin{tikzpicture}[>=latex,line join=bevel, font=\small, yscale=0.6,every path/.style={line width=0.5pt}, every node/.append style={line width=0.4pt}]
  \node[scale=1.5] (5) at (186.88bp,560.0bp) [draw,rectangle,rounded corners] {$\quad\ldots\quad$};
  \node (6) at (186.88bp,471.5bp) [draw,rectangle,rounded corners] {\cpp{if (!context && kadm5_init_krb5_context(&context))}};
  \node (33) at (148.88bp,164.0bp) [draw,rectangle,rounded corners] {\cpp{krb5_free_principal(context, *objp);}};
  \node (8) at (318.88bp,398.5bp) [draw,rectangle,rounded corners] {\cpp{switch (xdrs -> x_op)}};
  \node (34) at (245.88bp,91.0bp) [draw,rectangle,rounded corners,fill=yellow!40] {\textcolor{red}{/*<vul-start>*/}\cpp{break;}\textcolor{red}{/*<vul-end>*/}};
  \node (31) at (255.88bp,325.5bp) [draw,rectangle,rounded corners] {\cpp{case XDR_FREE:}};
  \node (32) at (249.88bp,252.5bp) [draw,rectangle,rounded corners] {\cpp{if (*objp != NULL)}};
  \node (36) at (394.88bp,18.0bp) [draw,rectangle,rounded corners] {\cpp{return TRUE;}};
  \definecolor{strokecol}{rgb}{0.0,0.0,0.0}
  \pgfsetstrokecolor{strokecol}
  \draw [black,dotted,->] (5) -- (6);
  \draw [pastelblue,->] (6) ..controls (67.184bp,442.7bp) and (46.876bp,426.07bp)  .. (46.876bp,399.5bp) .. controls (46.876bp,399.5bp) and (46.876bp,399.5bp)  .. (46.876bp,251.5bp) .. controls (46.876bp,223.13bp) and (70.105bp,202.14bp)  .. (33);
  \draw (66.751bp,325.5bp) node {context};
  \draw [pastelblue,->] (6) ..controls (140.94bp,440.7bp) and (122.88bp,422.07bp)  .. (122.88bp,399.5bp) .. controls (122.88bp,399.5bp) and (122.88bp,399.5bp)  .. (122.88bp,251.5bp) .. controls (122.88bp,231.37bp) and (129.41bp,209.68bp)  .. (33);
  \draw (148.38bp,325.5bp) node {\&context};
  \draw [pastelyellow,->] (6) ..controls (229.65bp,443.81bp) and (251.08bp,432.2bp)  .. (8);
  \draw [pastelred,->] (33) ..controls (272.97bp,168.41bp) and (279.88bp,166.57bp)  .. (279.88bp,164.0bp) .. controls (279.88bp,162.48bp) and (277.44bp,161.21bp)  .. (33);
  \draw [pastelred,->] (33) ..controls (184.66bp,136.81bp) and (199.91bp,125.64bp)  .. (34);
  \draw [pastelred,->] (8) ..controls (408.29bp,403.55bp) and (415.38bp,401.53bp)  .. (415.38bp,398.5bp) .. controls (415.38bp,396.75bp) and (413.01bp,395.33bp)  .. (8);
  \draw [pastelred,->] (8) ..controls (296.12bp,371.85bp) and (286.92bp,361.48bp)  .. (31);
  \draw [pastelyellow,->] (8) ..controls (333.33bp,361.2bp) and (340.98bp,330.44bp)  .. (328.88bp,307.5bp) .. controls (322.27bp,294.98bp) and (311.31bp,284.81bp)  .. (32);
  \draw [pastelyellow,->] (8) ..controls (365.97bp,368.19bp) and (385.88bp,350.15bp)  .. (385.88bp,326.5bp) .. controls (385.88bp,326.5bp) and (385.88bp,326.5bp)  .. (385.88bp,90.0bp) .. controls (385.88bp,75.938bp) and (385.88bp,60.31bp)  .. (36);
  \draw [pastelred,->] (34) ..controls (302.17bp,63.176bp) and (326.93bp,51.375bp)  .. (36);
  \draw [pastelred,->] (31) ..controls (253.78bp,299.64bp) and (252.99bp,290.39bp)  .. (32);
  \draw [pastelblue,->] (32) ..controls (211.13bp,228.79bp) and (202.52bp,222.78bp)  .. (195.13bp,216.5bp) .. controls (186.15bp,208.88bp) and (177.27bp,199.55bp)  .. (33);
  \draw (205.5bp,208.25bp) node[rotate=30] {*objp};
  \draw [pastelyellow,->] (32) ..controls (232.58bp,223.39bp) and (226.41bp,209.44bp)  .. (216.88bp,200.0bp) .. controls (212.26bp,195.43bp) and (206.95bp,191.4bp)  .. (33);
  \draw [pastelred,->] (32) ..controls (279.28bp,213.15bp) and (299.76bp,177.15bp)  .. (288.88bp,146.0bp) .. controls (285.28bp,135.7bp) and (278.8bp,125.89bp)  .. (34);
  \draw [pastelred,->] (32) ..controls (328.79bp,257.79bp) and (335.88bp,255.71bp)  .. (335.88bp,252.5bp) .. controls (335.88bp,250.65bp) and (333.51bp,249.17bp)  .. (32);
  \draw [pastelblue,->] (36) ..controls (456.69bp,23.823bp) and (464.0bp,21.636bp)  .. (464.0bp,18.0bp) .. controls (464.0bp,15.898bp) and (461.56bp,14.28bp)  .. (36);
  \draw (481.63bp,18.0bp) node {TRUE};
\end{tikzpicture}